\documentclass[]{ejm}
\usepackage{graphicx}
\usepackage{bm}
\usepackage{amssymb}
\usepackage{amsmath}
\usepackage{verbatim}

\newcommand\eb{\mathrm{e}}
\newcommand\mL{\mathcal{L}}
\newcommand\mJ{\mathcal{J}}
\newcommand\mN{\mathcal{N}}
\newcommand{\R}{{\mathbb R}}

\title{Additive noise effects in active nonlinear
spatially extended systems}

\author[M. Pradas et al.]{
M.\ns P\ls R\ls A\ls D\ls A\ls S\ls$\,^1$,\ns G.\ls A.\ns P\ls A\ls
V\ls L\ls I\ls O\ls T\ls I\ls S$\,^2$ S.\ns K\ls A\ls L\ls L\ls I\ls
A\ls D\ls A\ls S\ls I\ls S$\,^1$,\ns \break \ns D.\ls T.\ns P\ls
A\ls P\ls A\ls G\ls E\ls O\ls R\ls G\ls I\ls O\ls U$\,^2$,\ns \and
D.\ns T\ls S\ls E\ls L\ls U\ls I\ls K\ls O$\,^3$, }

\affiliation{$^1\,$Department of Chemical Engineering, Imperial College London, London, SW7 2AZ, UK \\
$^2\,$Department of Mathematics, Imperial College London, London, SW7 2AZ, UK\\
$^3\,$School of Mathematics, Loughborough University, Leicestershire, LE11 3TU, UK
}

\begin{document}
\label{firstpage}
\maketitle
\begin{abstract}
We examine the effects of pure additive noise on spatially extended
systems with quadratic nonlinearities. We develop a general
multiscale theory for such systems and apply it to the
Kuramoto-Sivashinsky equation as a case study. We first focus on a
regime close to the instability onset (primary bifurcation), where
the system can be described by a single dominant mode. We show
analytically that the resulting noise in the equation describing the
amplitude of the dominant mode largely depends on the nature of the
stochastic forcing. For a highly degenerate noise, in the sense that
it is acting on the first stable mode only, the amplitude equation
is dominated by a pure multiplicative noise, which in turn induces
the dominant mode to undergo several critical state transitions and
complex phenomena, including intermittency and stabilisation, as the
noise strength is increased. The intermittent behaviour is
characterised by a power-law probability density and the
corresponding critical exponent is calculated rigorously by making
use of the first-passage properties of the amplitude equation. On
the other hand, when the noise is acting on the whole subspace of
stable modes, the multiplicative noise is corrected by an
additive-like term, with the eventual loss of any stabilised state.
We also show that the stochastic forcing has no effect on the
dominant mode dynamics when it is acting on the second stable mode.
Finally, in a regime which is relatively far from the instability
onset, so that there are two unstable modes, we observe numerically
that when the noise is acting on the first stable mode, both
dominant modes show noise-induced complex phenomena similar to the
single-mode case.
\end{abstract}



\section{Introduction}
External or internal random fluctuations are ubiquitous in many
physical systems and can play a key role in their dynamics often
inducing a wide variety of complex spatio-temporal phenomena.
Examples can be found in several fields: From biology (such as
stochastic resonance accounting for the sensitivity that certain
species have, to weak but coherent signals in noisy environments)
and technological applications (e.g.~high-temperature
superconductivity~\cite{Bezrukov_1995,Wiesenfeld_1995}), to fluid
dynamics (e.g.~Rayleigh-B\'enard convection commonly used as a
prototype to study instabilities in systems out of
equilibrium~\cite{Cross.etal_1993}, contact line dynamics~\cite{Wylock_etal}  and 
waves in free-surface thin film flows~\cite{Kall2007,Pradas_pof}). Many of these natural phenomena and
technological applications can be described by model noisy spatially
extended systems (SES), i.e.~infinite-dimensional dynamical systems
described through stochastic partial differential equations (SPDEs)
with space-time dependence~\cite{Sagues.etal_2007}, such as the
noisy Swift-Hohenberg equation, or the noisy Kuramoto-Sivashinsky
(KS) equation. The spatio-temporal dynamics of noisy SES can be
dominated by many curious phenomena, such as noise-induced spatial
patterns~\cite{Garcia_Ojalvo.etal_1993} and noise-induced phase
transitions~\cite{Horsthemke_Lefever_1984}. Clearly, characterising
the influence of noise in SES is crucial for the understanding of
the inception and long-time complex spatio-temporal dynamics of
physical systems, as well as for the control and optimisation of
technological processes. The identification and understanding of
different regimes in parameter space, including the emergence of
underlying scaling laws is of particular interest in analytical and
computational studies of such systems.

One of the most widely studied complex phenomena induced by noise
effects is the transition between different observed system states
as the noise strength is continuously increased beyond a critical
value. A related question concerns the mechanisms by which
fluctuations in the system interact with nonlinearities to induce
ordered or regularised states (see
\cite{Sancho_book,Sagues.etal_2007}). It is now generally accepted
that one of the main mechanisms required to induce phase transitions
(e.g. between ordered-disordered phases), is the presence of pure
multiplicative noise\footnote{Multiplicative noise enters the system
through external effects, i.e. via the presence of noise on a
controlling parameter or through noisy fluctuations in the boundary
conditions; on the other hand, additive noise arises from
fluctuations of an internal origin such as thermal fluctuations,
which although they are present at the nanoscale, they can often
have macroscopic effects (e.g.~\cite{Mecke}).}, even though the
presence of combined multiplicative and additive noise has also been
shown to induce phase
transitions~\cite{Landa.etal_1998,Zaikin.etal_1999}. On the other
hand, complex phenomena such as stabilisation effects (but not phase
changes) can be induced by pure additive noise as reported in recent
numerical
investigations~\cite{Hutt_2008,Hutt.etal_2007,Obeid.etal_2010}.

In this study, we investigate the effects of pure additive noise on
unstable SES that are close to the primary bifurcation
(``instability onset"). A first stab at this problem was our recent
study in~\cite{Pavliotis_al2010} which focused on the influence of
highly degenerate noise, a particular type of additive noise, on
the KS equation -- the noisy KS equation is a paradigmatic model for
a wide spectrum of physical settings. In~\cite{Pavliotis_al2010} it
was shown that being close to the instability onset allows for a
global description of the KS equation in terms of a single dominant
mode. In the present work, by means of a multiple scale analysis for
general SES with quadratic nonlinearities and by appropriately
extending the singular perturbation methodology in~\cite{PavlSt08}
for such SES, we obtain an amplitude equation for the dominant mode
which then enables us to describe analytically the behaviour of this
mode and explore systematically the effects of noise. As an
application of the general methodology we examine the
noisy KS equation. We offer a derivation of this equation
for a hydrodynamic system, that of a thin-liquid film flowing over a
topographical substrate obtained asymptotically from governing
equations (Navier-Stokes) and wall and free-surface boundary
conditions.
In other contexts, the noisy KS equation  has
been adopted, for instance, as a model for sputtering processes used
to produce thin films in materials applications including
nanostructuring solid surfaces using ion beam erosion -- in
addition, a possible use of noise as a control tool is also
suggested and explored computationally, see
\cite{Cuerno_et_al95,Frost2003,
Lauritsen_et_al96,Christofides05,Christofides08}.

We observe that the dynamics, and in particular the statistical
properties of the fluctuating dominant mode, are largely dependent
on the nature of the applied noise and its strength. More precisely,
we consider a degenerate noise, in the sense that it is acting on
the subspace of stable modes alone, and scrutinise its effects on
the dynamics of the unstable mode. Several situations arise
depending on what stable mode (or modes) the noise is acting on. For
example, when the noise is acting on the first stable mode only, the
governing equation for the amplitude of the dominant (unstable) mode
is reduced to a purely temporal Stuart-Landau (SL) model
with pure multiplicative noise. The numerical simulations presented in our
previous study~\cite{Pavliotis_al2010} suggested that this type
of noise is in turn responsible for the different critical
transitions that the dominant mode undergoes as the noise strength
is increased, including an initial state of finite fluctuations, an
intermittent on-off state characterised by a burst-like dynamics,
and a completely stabilised state. It is noteworthy that such
critical transitions exhibit universal underlying scaling laws, in
the sense that the observed critical exponents are also ubiquitously
found in many different physical systems.
It is also noteworthy that such observed on-off intermittency is
actually a crucial feature of many nonlinear systems close to
criticality, and reflects a transition from order/coherence to a
disordered state.

Here, we rigorously derive the statistical properties of such
intermittency in terms of a universal law that does not depend on
the particular model under consideration. We further offer new
results and insights on the influence of noise on SES compared to
our previous study in~\cite{Pavliotis_al2010}. More specifically,
when the noise acts on the whole set of stable modes, we show that
the multiplicative term in the SL equation is corrected by an
additive-like term, and any stabilisation effect is eventually lost.
Additionally, we find that the noise can be filtered out by the
nonlinearities when it is acting on the second stable mode alone. In
this case, the dynamics of the dominant mode is not affected by the
stochastic forcing applied to the system. Our analytical findings
are in full agreement with numerical experiments of the KS equation,
thus providing a complete picture of the relevant effects of
additive noise on SES with quadratic nonlinearities close to the
instability onset.

Following the development of analytical and computational
descriptions of the phenomena near the primary bifurcation point, it
is important to consider the dynamics beyond criticality where the
multiscale analysis is no longer valid. To achieve this we revert to
numerical experiments and compute the resulting spatio-temporal
dynamics when
two linearly unstable modes are active.
Surprisingly, when we consider a highly degenerate
noise acting on the first stable mode alone, the
dynamics of the two dominant modes are found to follow
the same critical transitions as with the case close to the primary
bifurcation where only one unstable mode is present.

In \S \ref{Sec: Theoretical framework} we present the theoretical
framework for noisy SES close to the primary bifurcation. In \S
\ref{Sec: Amplitude Eq} we derive the generic amplitude equations
for the unstable mode components using multiscale techniques. Our
theory is then applied to the noisy KS equation in \S \ref{Sec: Case
study}, and both analytical and numerical results for the different
cases of noise considered are presented in \S \ref{Sec: Noise case
I}, where we take the noise acting on the first stable mode only, in
\S \ref{Sec: Noise case II}, where the noise acts on the first and
second stable modes, and in \S \ref{Sec: Noise case III}, where the
noise acts on the second stable mode only. A numerical study in a
regime relatively far from the instability onset is presented in \S
\ref{Sec: Far criticality}. Finally, our results are summarised
in~\S \ref{Sec: conclusions}.

\section{Theoretical framework: Noisy SES close to the
primary bifurcation} \label{Sec: Theoretical framework}

Typically, noisy SES can be described through SPDEs of the following
generic form:
\begin{equation}\label{Eq:General Eq}
\partial_t u=\mL u+
\mathcal{F}(u,\nabla u,\nabla^2 u,\ldots)+\tilde{\sigma}\xi(\bm{r},t),
\end{equation}
where $\mL$ is usually a linear  differential operator with constant
coefficients, and $\mathcal{F}$ is a nonlinear function of its
arguments. The field $u(\bm{r},t)$ describes the magnitude of a
quantity of interest in the system, and we also include the presence
of a stochastic additive forcing, given by the random variable
$\xi(\bm{r},t)$, with $\tilde{\sigma}$ its strength. The complexity of SES
of the form~(\ref{Eq:General Eq}) and their dynamics\footnote{It
should be emphasised, that already in the absence of noise ($\tilde{\sigma}
=0$), the presence of a wide range of spatial and temporal scales
coupled with each other through nonlinearities may give rise to a
complex dynamics such as transitions between different patterns and
spatio-temporal chaos~\cite{Cross.etal_1993}.} is such that it is
often quite difficult, if not impossible, to analyse them directly
either analytically or even numerically. It is therefore desirable
to obtain an appropriate low-dimensional description representing
the dynamics at a coarse-level and which can capture most, if not
all of the essential dynamic features of the original
high-dimensional model describing the actual problem.

Quite often in physical systems, the dominant nonlinearity is a
quadratic one whose functional form is dictated by simple symmetry
considerations. For example, in the context of free-surface
thin-film flows, the dominant nonlinearity is $u \partial_x u$,
associated with the interfacial kinematics due to mean
flow~\cite{Sergey} (the only other dominant nonlinearity, $u^2$, is
easily ruled out for systems whose spatial average is conserved,
i.e. $\partial_t \langle u \rangle_x = 0$). We then consider SES
with a quadratic nonlinearity, random forcing and a spatially
uniform rest state. The spatial domain is taken to be $[-L,L]^d$
where $d+1$ is the number of space and time dimensions. We note that even
though the multiscale techniques we shall be developing in this
study could, in principle be extended to multidimensional systems,
this extension is non-trivial and beyond the scope of the present
study (in fact, the theory of SPDEs in dimensions higher than one is
currently not well developed, at least for space-time white noise).
For this reason, we will consider Eq.~(\ref{Eq:General Eq}) in one
spatial dimension ($d=1$). We also assume that zero is an eigenvalue
of the linear operator $\mL$ with the corresponding eigenspace being
finite dimensional, and the remaining eigenvalues being in the left
complex half-plane.

Let us introduce a bifurcation parameter $\epsilon$, such that for
$\epsilon=0$ the system approaches its rest state as $t \to \infty$.
For $\epsilon> 0$, a bifurcation occurs leading to a finite number
of modes that are linearly unstable. Therefore, we assume that
$\mathcal{F}(u,\nabla u,\nabla^2 u,\ldots)=\epsilon^{2}\mJ
u+\mathcal{B}(u,u)$, where $\mJ $ and $\mathcal{B}(u,u)$ are a
linear differential operator with constant coefficients and a
quadratic nonlinearity (a symmetric bilinear map), respectively, so
that for $\epsilon\ll 1$, there appears at least one eigenvalue of
the perturbed operator $\mL+\epsilon^2\mJ $ in the right half-plane.
Equation~(\ref{Eq:General Eq}) is then written in the form:
\begin{equation}\label{Eq:Model}
\partial_t u=\mL u+\epsilon^{2}\mJ u+
\mathcal{B}(u,u)+\epsilon \sigma\xi(x,t).
\end{equation}
For the particular distinguished limit that we will consider in this
paper, and which leads to the most interesting coarse-grained
dynamics, we have rescaled the noise term in Eq.~(\ref{Eq:General Eq})
with $\tilde{\sigma}=\epsilon\sigma$ so that its
strength is of $O(\epsilon)$, in comparison to the $O(\epsilon^2)$
distance from criticality. Different distinguished limits, relevant
when different assumptions on the nonlinearity and the structure of
the noise are made, can also be considered, e.g.~\cite{Blomker2007, BlomkerMohammed2009}. The field $u$ can
then be projected onto the set of eigenfunctions
$\left\{\eb_k(x)\right\}$ for $k=1, \ldots ,\infty$ of the linear
operator $\mL$, i.e.~$u(x,t)=\sum_k \hat{u}_k(t)\eb_k(x)$. For
$0<\epsilon\ll 1$, the system is close to the bifurcation point and
is described by the presence of a finite number of unstable modes,
the ``dominant modes". In this regime, Eq.~(\ref{Eq:Model}) has two
widely separated time scales, corresponding to the (stable) fast and
(unstable) slow modes, allowing us to derive an amplitude equation
for the slow-dynamics of the dominant modes only, which belong to
the null space $\mN$ of the linear operator $\mL$.

\subsection{Amplitude equation reduction} \label{Sec: Amplitude Eq}
We start with the generic stochastic equation (\ref{Eq:Model}) on
the domain $[-L,L]$. For $\epsilon\ll 1$, the number of linearly
unstable modes is given by the dimensionality of $\mN$.
We are interested in the dynamics of the dominant modes  when
the stable modes, $\hat{u}_k(t)\eb_k$ for $\eb_k\in \mN^\perp$, are
randomly forced. Here, $\mN^\perp$ represents the subspace
of fast modes orthogonal to  $\mathcal{N}$. The noise term in
Eq.~(\ref{Eq:Model}) is therefore written as:
\begin{equation}\label{Eq:noise repr}
\xi(x,t)=\sum_{\mathrm{e}_k\in\mathcal{N}^{\bot}}q_k\dot{\beta}_k(t)\eb_k(x).
\end{equation}
%
The variable $\dot{\beta}_k(t)$ in the above equation
corresponds to uncorrelated white noise,
$\langle \dot{\beta}_m(t) \dot{\beta}_n(t')\rangle=\delta_{mn}\delta(t-t')$,
and $q_k$ represents the wave number dependence of the noise.
Considering now the behaviour of small
solutions  at time scales of $O(\epsilon^{-2})$, we define
$u(x,t)=\epsilon v(x,\epsilon^2 t)$ and use the scaling properties
of the white noise to transform Eq.~(\ref{Eq:Model}) to:
\begin{equation}\label{Eq:Model scaled}
\partial_t v=\epsilon^{-2}\mL v+\mJ v+\epsilon^{-1}\mathcal{B}(v,v)+
\epsilon^{-1} \sigma \xi(x,t).
\end{equation}
A detailed rigorous derivation of the pertinent amplitude equation for the
dominant mode in the case where $\mJ$ is the identity operator can
be found in Ref.~\cite{Blomker.etal_2007}. We shall extend this
previous formalism here for the general case of $\mJ $ being
any linear (differential) operator that commutes
with $\mL$, i.e.~both operators have the same eigenfunctions, and
for a finite dimensional kernel, i.e.~$\mathrm{dim}(\mathcal{N})=N_0$.

To obtain the amplitude equation for the dominant modes we consider a
finite dimensional truncation of the above system and keep
$M$ fast modes in the series expansion, so that the total number
of modes in the expansion is $N=N_0+M$.
We also consider Eq.~(\ref{Eq:Model scaled}) in a bounded domain, e.g. $[-\pi,\pi]$,
and we project the field $v$ onto  $\mathcal{N}$  to get $v_1=\mathcal{P}_c v$,  where $\mathcal{P}_c$
is the corresponding projector to the null space, and onto its
orthogonal subspace $\mathcal{N}^\perp$ to get $v_\perp=\mathcal{P}_sv$,
where $\mathcal{P}_s=\mathcal{I}-\mathcal{P}_c$ with $\mathcal{I}$
being the identity operator. The resulting system of equations
reads:
\begin{subequations}
\label{Eq:Scale sep}
\begin{eqnarray}
\partial_t v_1 & = &\mJ v_1+2\epsilon^{-1}\mathcal{P}_c \mathcal{B}(v_1,v_\perp)+
\epsilon^{-1}\mathcal{P}_c \mathcal{B}(v_\perp,v_\perp), \label{Eq:system a}{} \\
\partial_t v_\perp & = & (\mJ +\epsilon^{-2}\mL)v_\perp+
\epsilon^{-1}\mathcal{P}_s\mathcal{B}(v,v)+
\epsilon^{-1}\sigma \xi(x,t). \label{Eq:system b}
\end{eqnarray}
\end{subequations}
By applying now techniques from homogenisation
theory~\cite{PavlSt08} and assuming the condition for the bilinear map of
$\mathcal{P}_c\mathcal{B}(\eb_k,\eb_k)=0$ for $k=1,\dots,N$,
we can derive the homogenised (i.e. amplitude) equations that describe the dynamics of the unstable modes.
The detailed derivation for a multidimensional null space is given
in Appendix~\ref{App amplitude eq}.

In the case of 1D null space, we define the amplitude $A$ for the
dominant mode as $v_1=A(t)\eb_1$, and its corresponding equation is
given by the SL equation with multiplicative  noise
(see Appendix~\ref{App amplitude eq} for details):
\begin{equation}\label{Eq:SL}
\dot{A}=(j_1+\gamma_1\sigma^2) A-\gamma_2 A^3
+\sigma\sqrt{\gamma_a\sigma^2+\gamma_m A^2}\,\dot{W}(t),
\end{equation}
where $\dot{W}(t)$ is a white noise, and where the different
coefficients are given in Eq.~(\ref{Eq:coefficients SL App}) from
Appendix~\ref{App amplitude eq}.
We therefore see that the coefficients explicitly depend on the nature of the noise,
i.e., the values of $q_k$, and the properties of the quadratic nonlinearity, which
are dictated by $B_{n\ell k}$ given in Eq.~(\ref{Eq: B_nlk}).

In the case of a two-dimensional (2D) null space, the solution is
expanded as $v_1=a_1(t)\eb_1+a_2(t)\eb_2$, and the coupled amplitude equations we obtain
are of the form:
\begin{subequations}
\label{Eq:SL 2D}
\begin{eqnarray}
\dot{a}_1 & =&  (j_1+2\gamma_1 \sigma^2) a_1-\gamma_2 a_1A^2+2\sigma\sqrt{\gamma_a\sigma^2+\gamma_m A^2}\, \dot{W}_1(t), {} \\
\dot{a}_2 & =&  (j_1+2\gamma_1 \sigma^2) a_2-\gamma_2 a_2A^2+2\sigma\sqrt{\gamma_a\sigma^2+\gamma_m A^2}\, \dot{W}_2(t),
\end{eqnarray}
\end{subequations}
where we have defined the global amplitude $A(t)=\sqrt{a_1^2+a_2^2}$, and
the coefficients $\gamma_1$, $\gamma_2$,  $\gamma_a$, and $\gamma_m$ are given by
Eqs.~(\ref{Eq:coefficients SL App}).

\subsection{Analysis of the amplitude equation}

The stationary statistical properties of the amplitude dynamics
can be studied by solving the corresponding stationary Fokker-Planck
equation. In the 1D case, the stationary probability density function
(PDF) for the random variable $A$  in the Stratonovich interpretation
for natural boundary conditions is given by~\cite{Horsthemke_Lefever_1984,Mackey.etal_1990}:
\begin{equation}
 P(A)=\frac{N_c}{g(A)}\exp{\int^{A}\frac{2r(z)}{g^2(z)}}\mathrm{d}z,
\end{equation}
where $r(A)=(j_1+\gamma_1\sigma^2) A-\gamma_2 A^3$ and
$g(A)=\sigma\sqrt{\gamma_a\sigma^2+\gamma_m A^2}$, yielding in our
case

\begin{equation}\label{Eq:PDF 1D}
P(A)=N_c (A^2+\lambda^2\sigma^2)^{\alpha_1/2}\exp{(-\mu A^2)},
\end{equation}
with:
\begin{equation}\label{Eq:parameters}
\alpha_1(\sigma)=2\frac{j_1+(\gamma_1+\gamma_2\lambda^2)\sigma^2}{\gamma_m\sigma^2}-1,
\quad \mu(\sigma)=\frac{\gamma_2}{\gamma_m\sigma^2},
\end{equation}
%
where we have defined the parameter
$\lambda\equiv\gamma_a/\gamma_m$, and the normalisation constant $N_c$
is given as
$N_c=2\mu^{\frac{\alpha_1+1}{2}}/\Gamma_\lambda(\alpha_1)$, where
$\Gamma_\lambda(\alpha_1)=\int_{-\infty}^{\infty} x^{-1/2}(x+
\mu\lambda^2\sigma^2)^{\alpha_1/2}\eb^{-x}\:\mathrm{d}x$. It should be noted that
the statistical properties of the fluctuating dominant mode depend
on both the strength of the noise $\sigma$ and the values of $q_k$
(see \S \ref{Sec: Case study} below for the example of the KS
equation).

For a 2D null space, the stationary joint PDF for the two variables,
$G(a_1,a_2)$, can also similarly be obtained by computing the corresponding
stationary 2D Fokker-Planck equation. This yields to:
$$
G(a_1, a_2) \propto(a_1^2+a_2^2+\lambda^2)^{\alpha_1'/2}\exp{[-\mu'(a_1^2+a_2^2)]},
$$
where the parameters $\alpha_1'$ and $\mu'$ are obtained from the
expressions in Eq.~(\ref{Eq:parameters}) by replacing $\gamma_1$,
$\gamma_a$, and $\gamma_m$ with $2\gamma_1$, $4\gamma_a$, and
$4\gamma_m$, respectively. The interesting point now is to study the
behaviour of the PDF, $P(A)$, corresponding to the global amplitude
$A=\sqrt{a_1^2+a_2^2}$. To this
end, we move to a polar coordinate system $(A,\theta)$ by applying
the transformation $y_1=A\sin{\theta}$ and $z_1=A\cos{\theta}$, and
we impose the normalisation condition between both distributions,
$G(a_1,a_2)\mathrm{d}a_1\mathrm{d}a_2=P(A,\theta)\mathrm{d}A\mathrm{d}\theta$,
giving rise to a function of the form:
\begin{equation}
P(A)\propto A(A^2+\lambda^2)^{\alpha_1'/2}\exp{(-\mu' A^2)}. \label{Eq: PDF 2D}
\end{equation}
Let us now apply the theory outlined above on a prototype model
equation.

\section{Case study: the noisy Kuramoto-Sivashinsky equation}
\label{Sec: Case study}
Consider the noisy KS equation:
\begin{equation}\label{Eq:KS}
\partial_t u=-(\partial_x^2+\nu\partial_x^4)u-
u\partial_x u+\tilde{\sigma}\xi,
\end{equation}
normalised to $2\pi$-periodic domains so that $0<\nu=(\pi/L)^2$,
where $2L$ is the original length of the system. The equation
corresponds to an important class of SES, active-dissipative
nonlinear media, whose main features are the presence of mechanisms
for instability/energy production at long scales ($\partial_x^2 u$)
and stability/energy dissipation at short scales ($\partial_x^4 u$).
Both without and with the noise term, the KS equation has attracted
a lot of attention since it appears in a wide variety of
applications and physical phenomena. These include instabilities of
flame fronts, models of collisional trapped ion modes in plasmas,
dissipative turbulence, interfacial instabilities in two-phase
flows, reaction-diffusion systems, the control of surface roughness
in the growth of thin solid films by sputtering, step dynamics in
epitaxy, the growth of amorphous films, and models in population
dynamics~\cite{ion_beam_sputtering,Duprat.etal_2009,epitaxy,
Kura1976,Papageorgiou.etal_1990,Siva1977,Tseluiko.etal_2010}. A
detailed derivation of the noisy KS equation in the context of
thin-film hydrodynamics is given in  Appendix \ref{App noisy KS}.

The KS equation represents an ideal candidate for our studies due to the
extensive rigorous and computational results available. It is often used
as a paradigmatic partial differential equation (PDE) that has
low dimensional behaviour producing complex dynamics such as chaos -
\cite{HymNic86,HNZ86,JKT90,PapSmy91,SmyPap91,Wi02,WiHo99}.
%
The existence and
uniqueness  of solutions for the stochastic KS equation have been
proven in~\cite{DuanErvin}. The effect of weak additive white noise
on transitions between stable fixed-point solutions and stable
travelling-wave solutions of the KS equation has been considered
numerically in \cite{WZE} and the relationship between noise-induced
transitions and the underlying attractors are explored.

By assuming $\nu=1-\epsilon^2$ and $\tilde{\sigma}=\epsilon\sigma$,
Eq.~(\ref{Eq:KS}) is rewritten as:
\begin{equation}\label{Eq:KS scaled}
\partial_t u=-(\partial_x^2+\partial_x^4)u+\epsilon^2\partial_x^4u
-u\partial_x u+\epsilon\sigma\xi.
\end{equation}
We can therefore read off all the different terms in Eq.~(\ref{Eq:Model})
as $\mL=-\partial_x^2-\partial_x^4$, $\mJ =\partial_x^4$, and
$\mathcal{B}(u,u)=-u\partial_x u$. It is important to emphasise that in this
setting, the term $\epsilon^2\partial_x^4u$ represents a linear instability term that can
destabilize the dominant modes of the equation. Note that it is controlled by the
parameter $\epsilon^2$ which measures  the distance from bifurcation. In this sense,
a decrease in $\nu=1-\epsilon^2$ below the bifurcation point $\nu=1$ (and hence
increasing $\epsilon>0$)  reduces the linearly stabilizing term $\partial_x^4 u$ and
therefore destabilizes the dominant modes\footnote{Note
that when the KS equation is written in this form, multiscale analysis can be used in
order to obtain the amplitude equation in a rigorous and systematic way. See
Appendix A for details.}.
In our analysis we  assume solutions of zero mean and we shall
consider both the case of Dirichlet boundary conditions (DBC), i.e.~$u(-\pi,t)=u(\pi,t)=0$,
and periodic boundary conditions (PBC),  together with the additional condition that
the derivatives of $u$ are also periodic.
In the case of DBC we also consider an initial condition given as an odd periodic small
function, e.g.~$u(x,0)=\epsilon\sin{(kx)}$, and the null space of $\mL$ is 1D:
$$
\mN(\mL) = \mbox{span} \left\{\sin(\cdot)\right\},
$$
and the solution can be expanded in the basis
$\left\{\eb_k(x)=c_k\sin{(kqx)}\right\}$, where $q=\pi/L=1$ and
$c_k$'s are normalisation constants.
On the other hand, for PBC the null space of $\mL$ is 2D:
$$
\mN(\mL) = \mbox{span} \left\{\sin(\cdot), \, \cos( \cdot) \right\},
$$
and the solution is expanded in the exponential Fourier basis
$\{\eb_{2k-1}=c_{2k-1}\sin{(kqx)}$,
$\eb_{2k}=c_{2k}\cos{(kqx)}\}$  for  $k=1,2,\dots$.
By applying the multiscale formalism developed in \S \ref{Sec:
Amplitude Eq}, we then obtain that the amplitude equation for the
dominant mode is given either by Eq.~(\ref{Eq:SL}) for DBC or by
Eqs.~(\ref{Eq:SL 2D}) for PBC, with $j_1=1$ and the following
coefficients:
\begin{subequations}
\label{Eq:coeff KS}
\begin{align}
\gamma_1 &= -\frac38\frac{q_2^2}{\lambda_2(\lambda_2+\lambda_3)}+\frac18\sum_{n=3}^{N}\frac{q_n^2}{\lambda_n}
\biggl(\frac{n-1}{\lambda_n+\lambda_{n-1}}-\frac{n+1}{\lambda_n+\lambda_{n+1}}\biggr), {} \\
\gamma_2 &= \frac{1}{48}, \qquad \gamma_m =\frac{q_2^2}{576}, {} \\
\gamma_a &= \frac18\frac{q_2^2q_3^2}{\lambda_3(\lambda_2+\lambda_3)^2}+
\frac18\sum_{n=3}^{N}q_n^2\biggl(\frac{q_{n+1}^2}{\lambda_{n+1}(\lambda_n+\lambda_{n+1})^2}
+\frac{q_{n-1}^2}{\lambda_{n-1}(\lambda_n+\lambda_{n-1})^2}\biggl),
\end{align}
\end{subequations}
where $\lambda_k=k^4-k^2$, and the noise in the amplitude equation is
interpreted in the Stratonovich sense (see Appendix A). From the above expressions it is
interesting to note that several situations arise for different
cases of noise. For example, for a highly degenerate noise acting on
the first stable mode only (Case I, $q_k=\delta_{k,2}$), there is
only multiplicative noise ($\gamma_a=0$). On the other hand, when
the noise acts on both the second and third mode (Case II,
$q_k=\delta_{k,2}+\delta_{k,3}$), the multiplicative noise is
corrected by an additive-like term ($\gamma_a\ne 0$). Finally, when
the noise acts on the third mode only (Case III,
$q_k=\delta_{k,3}$), the resulting amplitude equation contains no
noise ($\gamma_m=\gamma_a=0$).
It is worth mentioning that in the case of a thin film flowing
down an uneven wall (see Appendix B), forcing the KS equation with a
highly degenerate noise would correspond to randomly vibrated
substrates with sinusoidal shape, and hence substrates with a
specific selected wavenumber, for example the second one (which
would correspond to Case I) or the third one (Case III).

In the following, we analyse analytically and numerically the
dynamics of the amplitude of the dominant mode $A(t)$ 
by considering the above three different cases of
noise separately. To solve numerically the KS equation we adopt a
pseudo-spectral method for the spatial derivatives that uses the
Fast Fourier Transform (FFT) to transform the solution to Fourier
space. The nonlinear terms are evaluated in real space and
transformed back to Fourier space by using the inverse FFT. The
solution is then propagated in time by making use of a fourth-order
Runge-Kutta scheme. In our simulations we have chosen a time step
of $\Delta t=0.1$, and the numerical error has been controlled by
monitoring some constant quantity, namely the spatially averaged
solution $\langle u \rangle=\frac{1}{2L}\int_{-L}^{L}u(x,t)\mathrm{d}x$,
so that with the chosen time step it remains constant during all
the numerical experiments.

\begin{figure}
\centering
\includegraphics[width=0.5\textwidth]{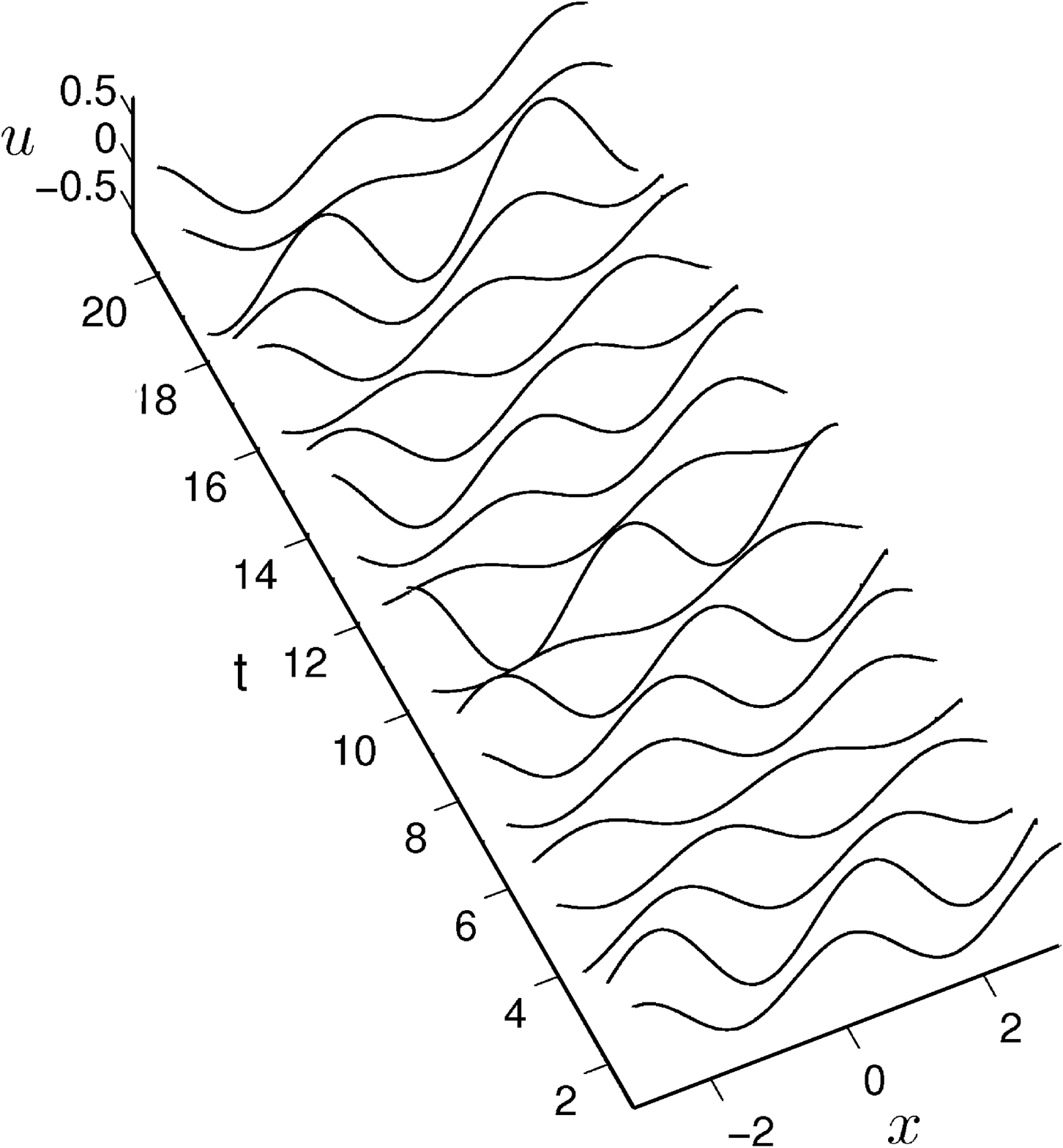}
\caption{Typical spatio-temporal evolution of the noisy KS
equation (\ref{Eq:KS}) with PBC for $\sigma=10$ and
$\epsilon=0.025$, and imposing a highly degenerate noise acting on
the first stable mode (Case I).}\label{Fig: spatio-temporal}
\end{figure}%

\subsection{Case I. Amplitude equation with multiplicative noise: Noise induced
critical state transitions}
\label{Sec: Noise case I}

We start by considering the case of highly degenerate noise acting
on the first stable mode only so that the resulting amplitude
equation for the dominant mode contains a pure multiplicative noise
term ($\gamma_a=0$). Typical snapshots of the spatio-temporal
evolution of the KS Eq.~(\ref{Eq:KS}) in this noise setting and
subject to PBC with $\sigma=10$ and $\epsilon=0.025$ are depicted in
Fig.~\ref{Fig: spatio-temporal}. The interesting point here is that
as a consequence of the multiplicative noise in the amplitude
equation, several state transitions may arise depending on the value
of the noise strength. As  was pointed out
in~\cite{Mackey.etal_1990}, the presence of multiplicative noise,
and in particular the fact that the coefficient
$\sigma\sqrt{\gamma_m}A$ from Eq.~(\ref{Eq:SL}) vanishes at $A=0$,
becomes crucial for the description of the asymptotic stability of
the stationary PDF in terms of Lyapunov functions, giving rise then
to different scenarios depending on the integrability of the PDF,
i.e.~whether the PDF can be normalised or not.
As we shall show now, different states can be observed by
simply changing the strength of the noise.
The results are presented separately for each case of
boundary conditions.
\begin{figure}
\centering
\includegraphics[width=0.7\textwidth]{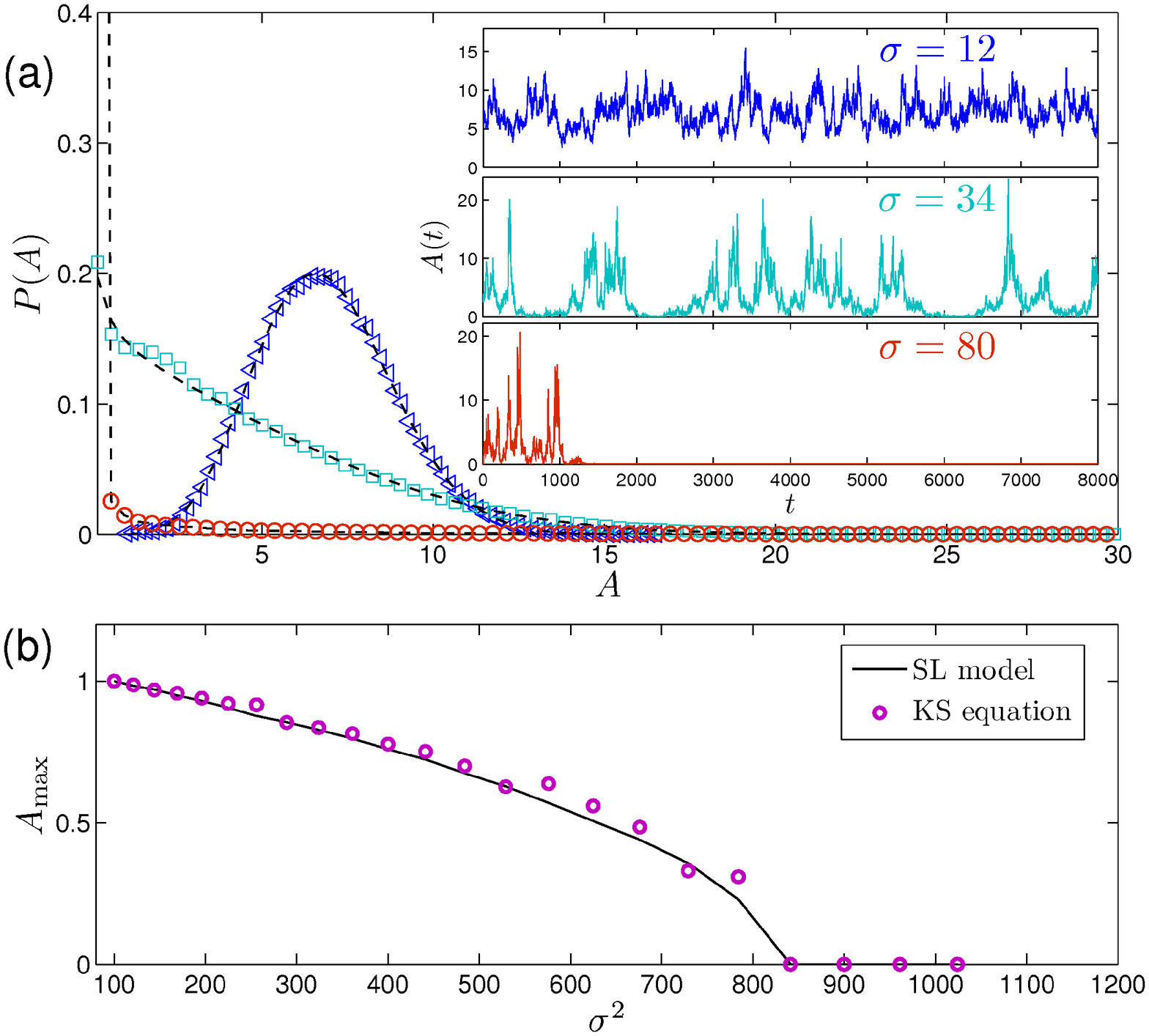}
\caption{(Colour online) Numerical results for the noisy KS
equation (\ref{Eq:KS}) integrated on a $[-\pi,\pi]$ domain with DBC
(1D null space). (a) PDF of the first-mode amplitude calculated  for
$\sigma=12$, $\sigma=34$, and $\sigma=80$, with $\epsilon=0.05$.
Dashed lines correspond to a data fit to a function of the form
given by Eq.~(\ref{Eq:PDF 1D}) with $\lambda=0$. The inset shows the
typical fluctuations of the amplitude at  each of the three values
of $\sigma$. (b) Maximum location of the amplitude PDF at different
values of $\sigma$. The solid line corresponds to the numerical
solution of the SL model. All values have been normalised to the
corresponding value at $\sigma=10$.}\label{Fig: 1D highly}
\end{figure}%

\begin{figure}
\centering
\includegraphics[width=0.95\textwidth]{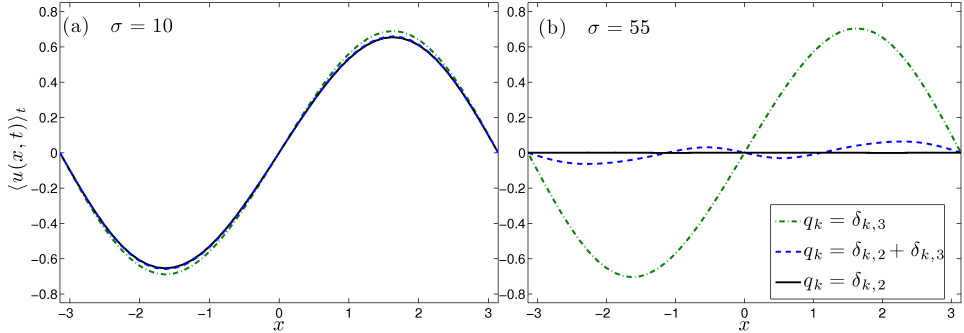}
\caption{(Colour online) Time averaged solution of the noisy KS equation
solved with DBC for $\epsilon=0.1$ and $\sigma=10$ (a) and $55$ (b) for
the case of: highly degenerate noise acting on the second mode (case I, solid line),
degenerate noise acting on both the second and third mode (case II, dashed line),  and
highly degenerate noise acting on the third mode (case III, dot-dashed line).}
\label{Fig: averaged sol}
\end{figure}

In the case of DBC, the PDF of the dominant mode amplitude,
$A(t)= \hat{u}_1(t)$, is given by Eq.~(\ref{Eq:PDF 1D})
with $\lambda=0$. In this case there exist different states for the
fluctuating amplitude that can be characterised by looking both at the
maximum location and integrability of the amplitude
PDF~\cite{Horsthemke_Lefever_1984,Mackey.etal_1990}.
First, we observe that as long as $\alpha_1>0$ the maximum of $P(A)$ occurs at
a finite value, $A_\mathrm{max}>0$, and then the state of $A$ is
characterised by finite fluctuations around a mean value (state I).
On the other hand, for $-1<\alpha_1\leq 0$, the maximum is located
at zero, $A_\mathrm{max}=0$, and the amplitude fluctuates
intermittently between zero and a finite value (state II, see also
below in \S \ref{Sec:Universal} for a detailed statistical analysis
of this state). These two states are separated by the critical
value:
\begin{equation}\label{Eq:sigma I}
\sigma_\mathrm{I}=(\gamma_m/2-\gamma_1)^{-1/2}.
\end{equation}
Note that for $\gamma_1>0$, this transition can only be observed as
long as $\gamma_m>2\gamma_1$, while it is always observed for
$\gamma_1<0$.  By computing the maximum of $P(A)$ at different
values of $\sigma$ we can then characterise the critical behaviour
as: $A_\mathrm{max}=\vert\sigma_\mathrm{I}^2-
\sigma^2\vert^{1/2}/(\sigma_\mathrm{I}\sqrt{\gamma_2})$ for $\sigma
\leq \sigma_\mathrm{I}$, and $A_\mathrm{max}=0$ otherwise, so that
$A_\mathrm{max}$ and $\sigma^2$ are the order and control parameter,
respectively, describing the critical transition. By using now the
particular values obtained for the KS equation [from
Eq.~(\ref{Eq:coeff KS}) we get $\gamma_1=-1/2688$, $\gamma_2=1/48$,
and $\gamma_m=1/576$] we find $\sigma_\mathrm{I}=28.3$ in excellent
agreement with the numerical results (presented in Fig.~\ref{Fig: 1D
highly}, where we obtain $\sigma_\mathrm{I}\simeq 29$). Finally, we
note that since $\gamma_1<0$, a second transition occurs when
$\alpha_1 \leq-1$. In this situation, it is not possible
to define Lyapunov functions to show the existence of a globally stable
stationary PDF for $A$ which can be normalised~\cite{Mackey.etal_1990}.
As a consequence, the PDF converges to  a Dirac delta function, $P(A)=\delta(A)$,
describing then a completely stabilised state with $A=0$ (state III). The critical
value $\sigma_\mathrm{II}$ for this second transition corresponds
to:
\begin{equation}\label{Eq:sigma II}
\sigma_\mathrm{II}=\sqrt{1/\vert\gamma_1\vert},
\end{equation}
which for the KS equation gives $\sigma_\mathrm{II}=51.8$, in very
good agreement with the numerical results [see the bottom panel in
the inlet of Fig.~\ref{Fig: 1D highly}($a$)]. Interestingly,  such
stabilisation effect is reflected on the time averaged solution
$\langle u(x,t)\rangle_t$ [see
solid line in Fig.~\ref{Fig: averaged sol}(b)]: since the dominant
mode has been completely stabilised, the dynamics of the solution is
only affected by the noise coming from the stable modes, which is wiped out
after time averaging, obtaining therefore a zero spatial solution,
i.e.~$\langle u(x,t)\rangle_t=0$. This is in contrast to the
inhomogeneous solution observed for smaller values of $\sigma$
corresponding to state I [see solid line in Fig.~\ref{Fig: averaged
sol}(a)].

When we consider the case of PBC,
the amplitude PDF is given by
\begin{equation}\label{Eq:P_A 2D case I}
 P(A)\propto A^{\alpha_2}\exp{(-\mu'A^2)},
\end{equation}
where now $\alpha_2=(1+2\gamma_1\sigma^2)/(2\gamma_m\sigma^2)$ and
$\mu'=\gamma_2/4\gamma_m$.
We first note that the transitions between different states can only
occur if and only if $\gamma_1 <0$. In such a case, the critical values for the
first and second transitions are found to be:
\begin{equation}\label{Eq:sigma crit 2D}
\sigma_\mathrm{I}=\sqrt{1/2\vert\gamma_1\vert},\qquad
\sigma_\mathrm{II}=[2(\vert\gamma_1\vert-\gamma_m)]^{-1/2}.
\end{equation}
Noteworthy is that the second transition can only occur as long as
$\gamma_m<\vert\gamma_1\vert$. Otherwise, the completely stabilised
state III is never observed, and the PDF tends to $P(A)\sim
A^{\alpha_\infty}$ as $\sigma\to\infty$, with
$\alpha_\infty=-\vert\gamma_1\vert/\gamma_m$. By using the KS
coefficients given in Eqs.~(\ref{Eq:coeff KS}), we obtain the first
transition to occur at $\sigma_\mathrm{I}=36.3$, in excellent
agreement with the numerical results presented in Fig.~\ref{Fig: 2D
highly}. The second transition, however, cannot be observed since
the condition $\gamma_m<\vert\gamma_1\vert$ does not hold, giving an
asymptotic behaviour with $\alpha_\infty=-0.21$, again in very good
agreement with the numerical data [cf.~Fig.~\ref{Fig: 2D
highly}($b$)].

It is important to remark that the reason why we observe such
large values of $\sigma$ for which the different transitions occur
(see e.g.~Fig.~\ref{Fig: 1D highly} with $\sigma=12, 34, 80$) is a
consequence of the scaling of the noise we used in
Eq.~(\ref{Eq:Model}). The actual noise we are considering  is
$\mathcal{O}(\epsilon\sigma)$, leading therefore to values which are
much smaller, namely $0.6, 1.7$, and $4$, respectively, for the
noise strength used in Fig.~\ref{Fig: 1D highly}. Let us also
emphasize that strictly speaking, such scaled perturbations are
still beyond the  region of validity of the asymptotic analysis,
$\epsilon \rightarrow 0$ for fixed $\sigma$,
i.e.~$\epsilon\sigma=4>1$. Nevertheless, we have performed extensive
numerical computations by choosing different values of $\epsilon$
ranging in $\epsilon\in [0.02,0.1]$, observing the same results in
all the cases, which then points to the robustness of the
perturbation scheme. This also indicates that our numerical results
are consistent with the asymptotic analysis presented in \S
\ref{Sec: Amplitude Eq}.

\begin{figure}
\centering
\includegraphics[width=0.68\textwidth]{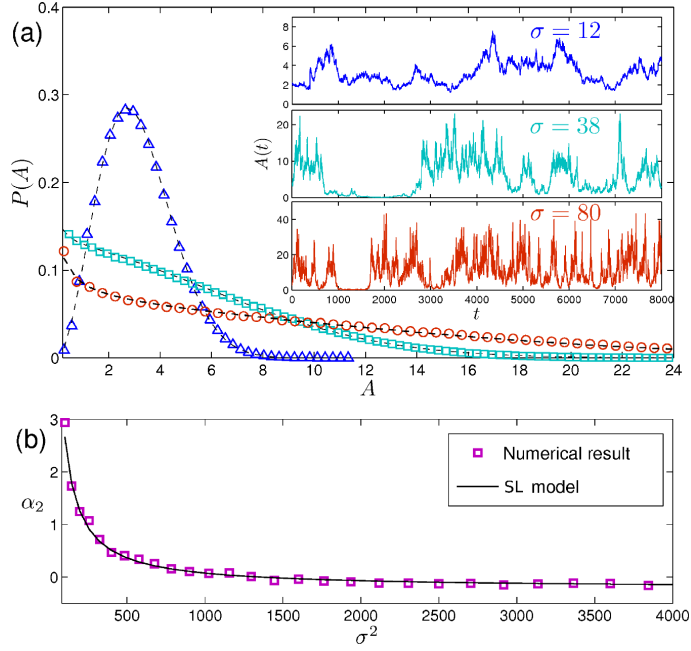}
\caption{(Colour online) Numerical results for the noisy KS
equation (\ref{Eq:KS}) integrated on a $[-\pi,\pi]$ domain with
PBC (2D null space). (a) PDF of the
first-mode amplitude calculated  for $\sigma=12$,
$\sigma=38$, and $\sigma=80$, with
$\epsilon=0.025$. Dashed lines correspond to a data fit to a function of the
form given by Eq.~(\ref{Eq:P_A 2D case I}). The inset shows the typical
fluctuations of the amplitude at  each of the three values of
$\sigma$. (b) Exponent $\alpha_2$ obtained from the data fit at different values of $\sigma$.
The solid line corresponds to the analytical solution of the
SL model given by Eq.~(\ref{Eq:P_A 2D case I}).}\label{Fig: 2D highly}
\end{figure}%

\subsubsection{Universal intermittent behaviour}\label{Sec:Universal}
Let us focus here on the intermittent behaviour observed between the
two transitions, $\sigma_\mathrm{I}<\sigma<\sigma_\mathrm{II}$, for
DBC. It is worth emphasising at this point that on-off intermittency
is a crucial universal feature of many nonlinear systems close to
criticality, and reflects a transition from order/coherence to a
disordered state (hence understanding the statistical properties of
intermittency is crucial for the characterisation of this
transition). In our case, it reflects the transition between an
initially inhomogeneous state in space to a final zero state (see
solid line in Fig.~\ref{Fig: averaged sol}).

Figure \ref{Fig: intermittency} depicts the dynamics of the
amplitude calculated by using $\sigma=48$, which for DBC is close to
the second critical transition $\sigma_\mathrm{II}$. In this regime,
fluctuations are clearly dominated by an on-off intermittent, or
burst-like behaviour. As was pointed out in
Refs.~\cite{Heagy.etal_1994,Platt.etal_1993}, such kind of on-off
intermittency can be characterised by studying the statistical
properties of the waiting time between two consecutive bursts,
defined as large fluctuations above a given small threshold,
$A>c_\mathrm{th}$. Given the SL amplitude equation (\ref{Eq:SL}), we
can obtain an analytic expression for the PDF of the waiting times,
$P(T)$, as follows.
\begin{figure}
\centering
\includegraphics[width=0.9\textwidth]{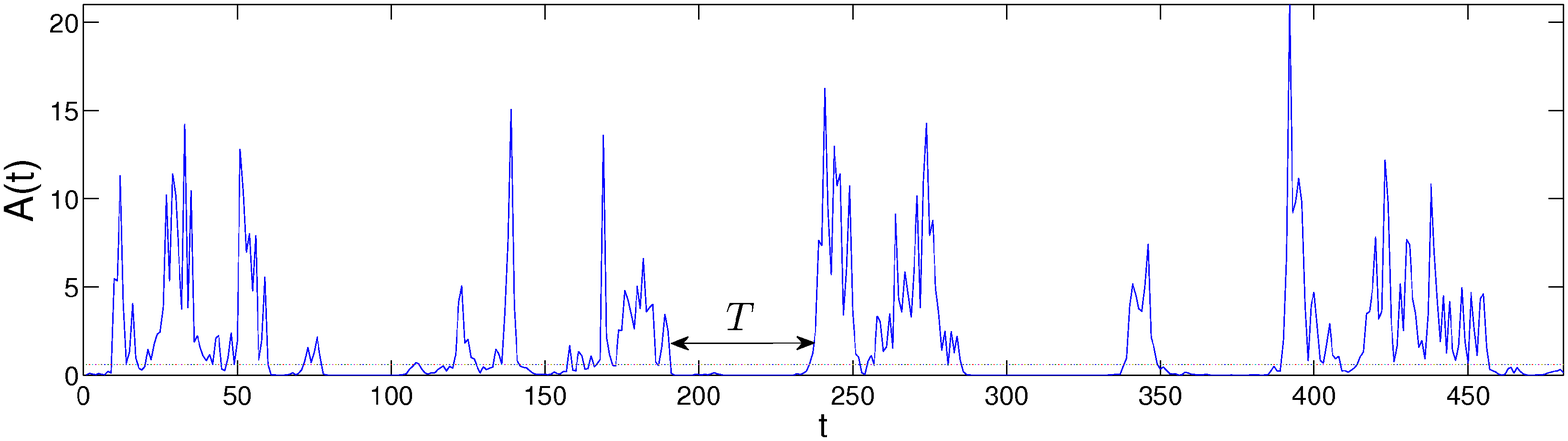}
\caption{(Colour online) Time evolution of the dominant mode
amplitude $A(t)$ for DBC with $\sigma=48$.  The waiting time between
two consecutive large events is denoted as $T$. The dashed line
represents the position of the chosen threshold $c_\mathrm{th}$
which defines the zero state as $A\le c_\mathrm{th}$.}\label{Fig:
intermittency}
\end{figure}%
\begin{figure}
\centering
\includegraphics[width=0.9\textwidth]{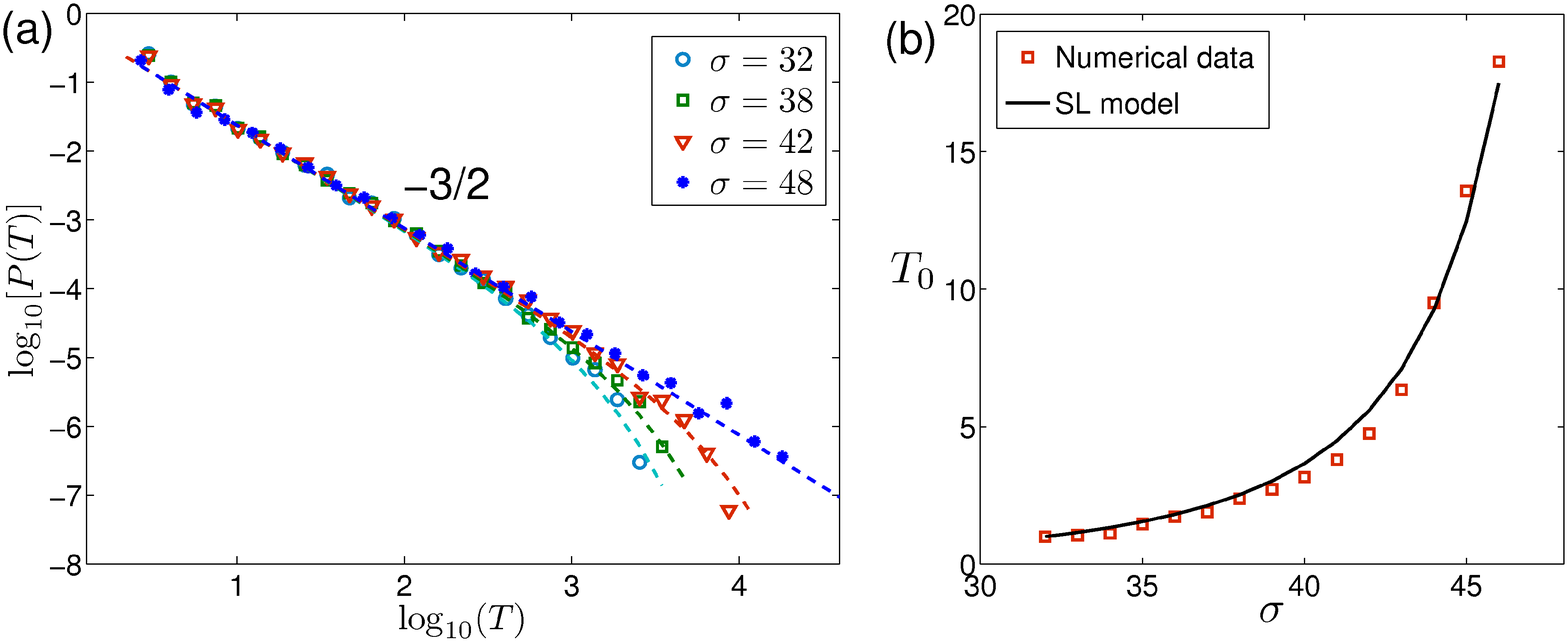}
\caption{(Colour online) Statistical analysis of the waiting times
between two consecutive bursts observed in the intermittent state II
for DBC. (a) PDF of the waiting times calculated by using different
noise strengths. The dashed line corresponds to a data fit to
$P(T)=N T^{-\tau}\exp{(-T/T_0)}$ with $\tau=1.5$. The value of the
threshold used to define the waiting times $T$ is $c_\mathrm{th}=1$.
(b) Value of the fitted time scale $T_0$ as a function of $\sigma$.
The solid line corresponds to the analytical expression given in
Eq.~(\ref{Eq:P(T)}). All values have been normalised to the
corresponding value at $\sigma=12$.} \label{Fig: waiting T 1 unst}
\end{figure}%

Let us assume an initial value which is below the threshold, i.e.,
$A_i\equiv A(t=0)<c_\mathrm{th}$. We then seek the probability
$P(T)$ that at time $T$, the variable $A$ reaches the threshold
$c_\mathrm{th}$  for the first time. In this close-to-zero state,
the amplitude is found numerically to be $A\lesssim 0.1$, and
therefore small enough to neglect the nonlinear term in the SL
equation. More precisely, we first introduce the transformation
$y=\log{A}$, and assume that $y\ll 0$ to obtain the following
linearised equation:
%
\begin{equation}
\dot{y}=\kappa+\sigma\sqrt{\gamma_m}\,\dot{W}(t),
\end{equation}
defined in the semi-infinite domain $y\in (-\infty,y_0]$ with
adsorbing boundary conditions at the origin
$y_0\equiv\log{c_\mathrm{th}}$, and where
$\kappa=1+\gamma_1\sigma^2$. We note that the choice of an adsorbing
boundary condition at the origin is required to ensure that the
variable $A$ leaves the domain once it reaches the threshold. The
problem now reduces to finding the probability that at time $T$ the
new variable $y$ reaches the origin for the first time. This
corresponds to the well-known ``first-passage probability" (FPP) of
the random walk in semi-infinite domains (see
Ref.~\cite{FirstPassage_book}), which simply reduces to solving the
Fokker-Planck equation for the probability $p(y,t)$:
\begin{equation}
 \partial_t p(y,t)=-\kappa\partial_y p(y,t)+\gamma_m\sigma^2\frac12 \partial_y^2
 p(y,t). \nonumber
\end{equation}
This can readily be done by using the method of images, obtaining:
%
\[ p(y,t;y_i,0)=\frac{1}{\sqrt{\pi4\gamma_m\sigma^2 t}}[\mathrm{e}^{-(y-y_i+\kappa t)^2/4\gamma_m\sigma^2 t}-
\mathrm{e}^{\kappa y_i/(\gamma_3\sigma)^2}\mathrm{e}^{-(y+y_i+\kappa t)^2/4\gamma_m\sigma^2 t}], \]
%
where $y_i=\log{A_i}$ is the initial value and we have assumed without loss
of generality that $c_\mathrm{th}=1$,
and  therefore $y_0=0$. The FPP can now be obtained as:
\begin{equation}
 P(T)=-\int_{-\infty}^{0}\partial_t p(y,t;y_i,0)\vert_{t=T}\:\mathrm{d}y, \nonumber
\end{equation}
from which we get:
\begin{equation}
P(T) =\frac{-y_i}{\sqrt{\pi4\gamma_m\sigma^2 T^3}}\mathrm{e}^{-(-y_i+
\kappa T)^2/4\gamma_m\sigma^2 T}.\nonumber
\end{equation}
In the long-time limit $T\to\infty$, the above expression becomes:
\begin{equation}\label{Eq:P(T)}
 P(T)\approx T^{-3/2}\mathrm{e}^{-T/T_0},
\end{equation}
with a time scale
$T_0=[2\sigma\sqrt{\gamma_m}/(1+\gamma_1\sigma^2)]^2$. Therefore, at
the critical point $\sigma=\sigma_\mathrm{II}$ the above PDF becomes
a pure power-law dictated by a universal exponent $3/2$ that does
not depend on the particular model we are using, i.e., it does not
depend on the coefficients of the SL equation. The numerical results
obtained with the KS equation are presented in Fig.~\ref{Fig:
waiting T 1 unst} where we can see an excellent agreement with the
above expression. It should be noted that far from the critical
point ($\sigma<\sigma_\mathrm{II}$), the power-law is exponentially
corrected with the time scale $T_0$ [cf.~Fig.~\ref{Fig: waiting T 1
unst}($b$)].

\subsection{Case II. Amplitude equation with both additive and multiplicative noise}
\label{Sec: Noise case II}
\begin{figure}
\centering
\includegraphics[width=0.68\textwidth]{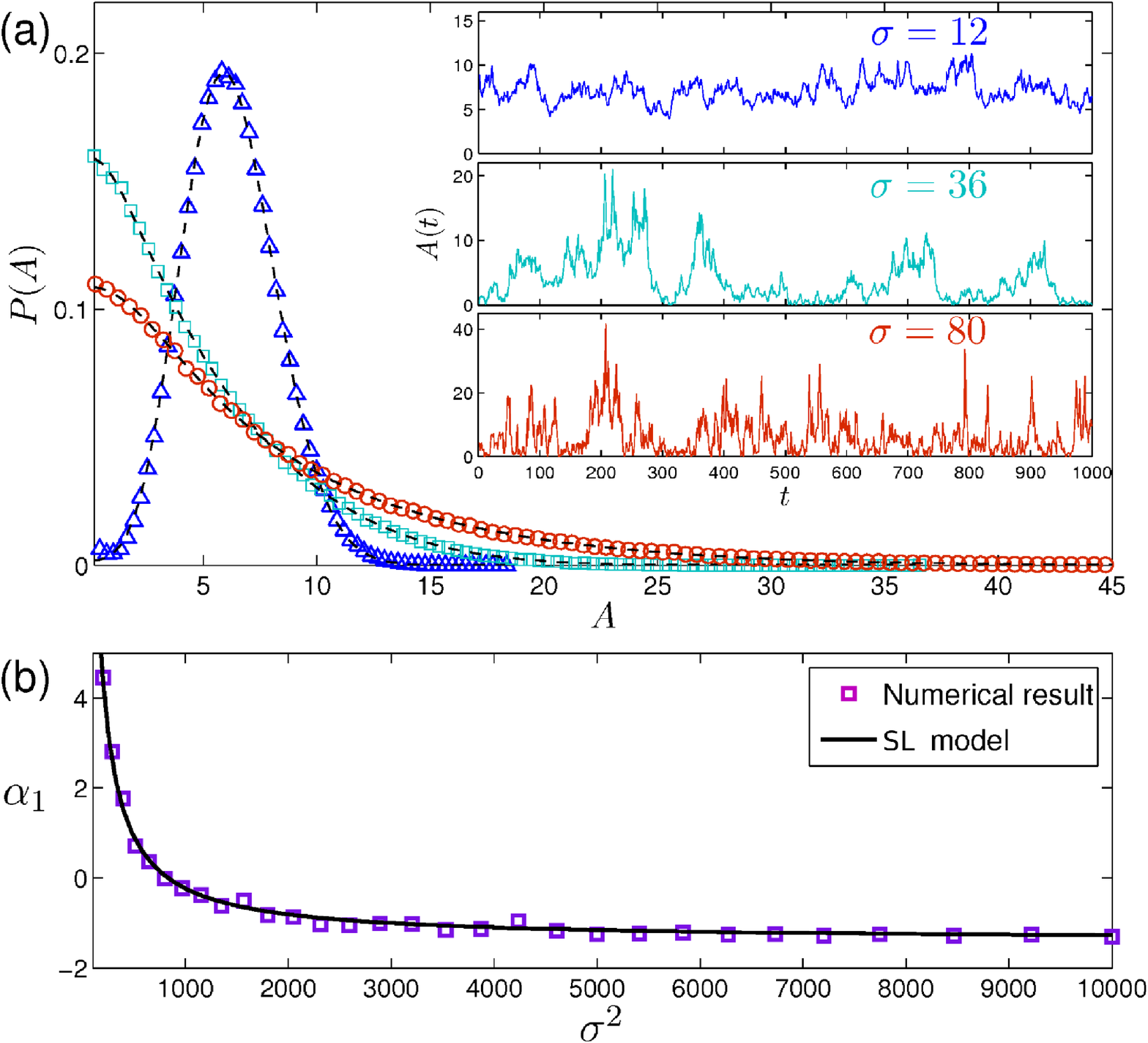}
\caption{(Colour online) Numerical results for the noisy KS
equation (\ref{Eq:KS}) integrated on a $[-\pi,\pi]$ domain with DBC and with
a noise acting on the second and third stable modes (case II). (a) PDF of the
first-mode amplitude calculated  for $\sigma=12$,
$\sigma=36$, and $\sigma=80$, with
$\epsilon=0.1$. The inset shows the typical
fluctuations of the amplitude at  each of the three values of
$\sigma$. (b) Exponent $\alpha_1$ obtained from the PDF  at different values of $\sigma$.
The solid line corresponds to the analytical solution of the
SL model given by Eq.~(\ref{Eq:parameters}).}\label{Fig: KS 1D noise on 2 3}
\end{figure}%

Let us now consider the noise term acting on both the second and
third mode, i.e.~$q_k=\delta_{k,2}+\delta_{k,3}$ with DBC. In this
setting, the amplitude equation is given by Eq.~(\ref{Eq:SL}) with
both $\gamma_a$ and $\gamma_m$ being different from zero. Although
the noise is still highly degenerate, it induces the same effect as
when we consider the noise acting on the whole subspace of stable
modes, i.e., $q_k=1$ for $k\ge 2$ (and therefore having space-time
white noise). For simplicity, we shall therefore restrict our
analysis to the case with only the third and second modes being
stochastically forced.

First, we note that in this case of noise we can still identify two
different dynamic behaviours (states) for the dominant amplitude. At
low values of $\sigma$ we have that, as it occurred in case I, the
location of the maximum of $P(A)$ is found to be at a finite value,
$A_\mathrm{max}>0$, and the dominant mode fluctuates around a finite
mean value. As the noise strength is increased, the maximum location
approaches zero and the first mode component $\hat{u}_1$ may reach
zero. Computing the maximum location by considering the PDF given by
Eq.~(\ref{Eq:SL}) we obtain that the critical value,
$\sigma_\mathrm{I}$, separating these two states corresponds to the
same value as in case I of noise given by Eq.~(\ref{Eq:sigma I}). In
this situation, however, the presence of the additive noise keeps
the first mode from remaining at the zero position, and hence ruling
out the intermittent dynamics observed in the previous section.
Further, as the noise strength is increased, the PDF can now be
always normalised so that the completely stabilised state III is
never observed.

Figure \ref{Fig: KS 1D noise on 2 3} depicts typical numerical
results for the KS equation. The coefficients of the SL equation in
this case are $\gamma_1=-1/2688+1/52416$, $\gamma_2=1/48$,
$\gamma_a=1/580608$, and $\gamma_m=1/576$, so that we have
$\sigma_\mathrm{I}\simeq 28.6$. To compare with the results presented
in \S~\ref{Sec: Noise case I}, we compute the amplitude of the
dominant mode by taking the absolute value of the first mode,
i.e.~$A(t)=\vert \hat{u}_1(t)\vert$, for different strengths
of the noise. The PDFs of $A$ for $\sigma=12$, $36$ and $80$ are
presented in Fig.~\ref{Fig: KS 1D noise on 2 3}($a$). We can see
that the presence of the extra term controlled by $\gamma_a$ in the
amplitude equation keeps the dominant mode away from zero and hence
of being completely stabilised.  As before, we find excellent
agreement between the numerical results and the analytical
derivation given by Eq.~(\ref{Eq:SL}). Indeed, Fig.~\ref{Fig: KS 1D
noise on 2 3}($b$) shows a comparison of the numerical value of the
exponent $\alpha_1$ obtained from a data fit of the PDF with the
analytical value given by Eq.~(\ref{Eq:parameters}) for different
strength of the noise. Moreover, by computing the time average of
the spatio-temporal solution we can see that at low values of the
noise strength ($\sigma=10$), the averaged solution still retains
the sinusoidal shape [see dashed line in Fig.~\ref{Fig:
spatio-temporal}($a$)], corresponding to the first state where
$A_\mathrm{max}>0$. At higher noise strength values ($\sigma=55$),
however, the averaged solution losses the sinusoidal shape [see
dashed line in Fig.~\ref{Fig: spatio-temporal}($b$)], corresponding
to the second state with $A_\mathrm{max}=0$. In contrast to the
stabilised state observed in case I of noise, the averaged solution
remains now noisy since the dominant mode is never completely
stabilised.

\subsection{Case III. Amplitude equation without noise}
\label{Sec: Noise case III} Finally, we consider the case of DBC
with a noise acting only on the third mode ($q_k=\delta_{3k}$). In
this setting, both noise coefficients are zero
($\gamma_m=\gamma_a=0$) with a positive coefficient $\gamma_1>0$ for
any value of the disorder strength. As a result, the dynamics of the
dominant mode is not affected by the noise, and only its steady
state $A_{st}$, given by the stationary solution of the
deterministic SL Eq.~(\ref{Eq:SL}) as:
\begin{equation}
 A_{st}=\sqrt{\frac{1+\gamma_1\sigma^2}{\gamma_2}},
\end{equation}
will be affected for sufficiently large $\sigma$. When we apply this
noise setting to the KS equation we obtain $\gamma_1=1/52416$, so
that only for values of noise strength as large as $\sigma\sim 230$
such effect will start to be relevant. As we see in Fig.~\ref{Fig:
averaged sol} (see dot-dashed line), the time averaged solution
$\langle u(x,t)\rangle$ is not affected when the noise is increased
up to $\sigma=55$. It is also worth emphasising that we would obtain
the same result had we considered a noise acting on all the stable
odd modes. It means that this kind of noise acting on the odd modes only
is filtered out by the quadratic nonlinearity interaction, such that it
has no effect on the dominant mode dynamics.

\section{Numerical results in a regime beyond the instability onset}
\label{Sec: Far criticality}

\begin{figure}
\centering
\includegraphics[width=0.95\textwidth]{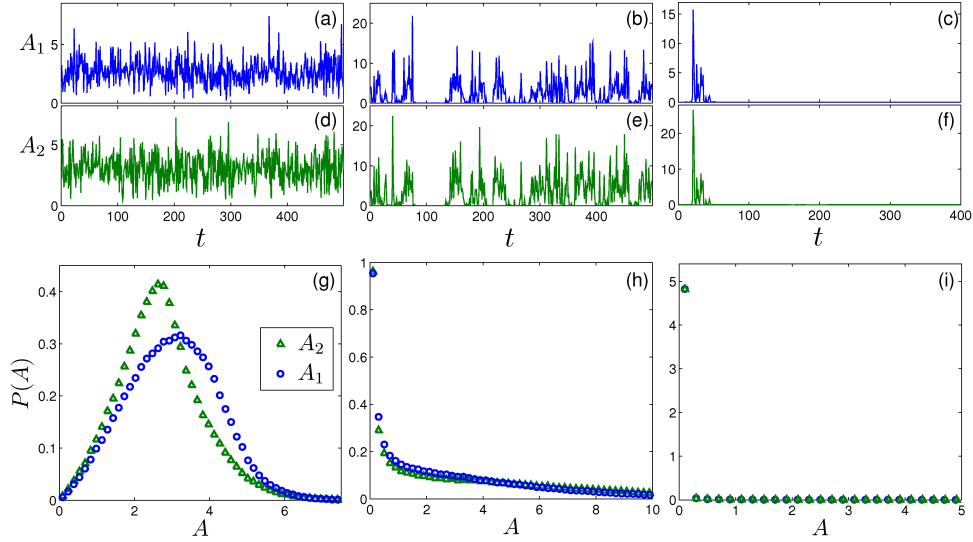}
\caption{(Colour online) Numerical results for the noisy KS
equation (\ref{Eq:KS}) integrated on a $[-\pi,\pi]$ domain in a regime
far from the stability onset ($\epsilon=0.87$) with a highly
degenerate noise acting on the first stable mode. Panels (a), (b),
and (c) show the time signal of the amplitude of the first dominant
mode $A_1(t)$ for $\sigma=40$, $230$ and $270$, respectively. Panels
(d), (e), and (f), show the corresponding time series of the second
unstable mode $A_2(t)$ for the same noise strengths. Bottom panels
(g), (h), and (i) show the PDF of each mode for the three different
values of $\sigma$.}\label{Fig: 2 unst}
\end{figure}%
Let us consider now the situation where there are more than one
unstable modes. Clearly, the theory presented in \S \ref{Sec:
Amplitude Eq} is not expected to be valid in this case. We shall
therefore numerically integrate the noisy KS equation by taking
$\epsilon=0.87$ such that the number of unstable modes is two. We
are interested to see whether stabilisation phenomena as presented
in \S \ref{Sec: Noise case I} for noise case I can still be observed
by appropriately tuning the noise. To this end we choose the noise
so that it acts on the first stable mode, corresponding to taking
$q_k=\delta_{k,3}$, and we study the dynamics of the amplitude of
both the first ($A_1(t)=\vert \hat{u}_1(t)\vert$) and second
($A_2(t)=\vert \hat{u}_2(t)\vert$) modes.

\begin{figure}
\centering
\includegraphics[width=0.9\textwidth]{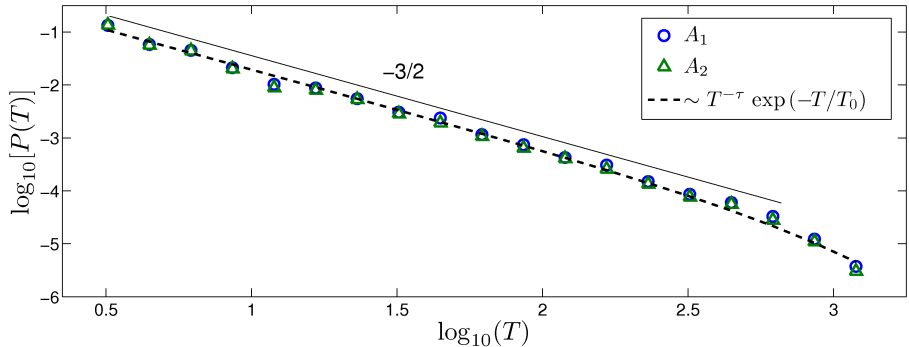}
\caption{(Colour online) PDF of the  waiting times between
two consecutive bursts obtained from the time series of
the two dominant modes $A_1$ and $A_2$ for
$\sigma=260$. The dashed line corresponds to a data fit to
$P(T)=N T^{-\tau}\exp{(-T/T_0)}$, with $\tau=1.5$.}\label{Fig: waiting T 2 unst}
\end{figure}

The corresponding numerical results are presented in Fig.~\ref{Fig:
2 unst}. Noteworthy is that both modes can be completely stabilised
as the noise strength is increased up to values of $\sigma\sim 270$,
going through the same critical transitions as in \S \ref{Sec: Noise
case I}. As before, we perform a statistical analysis of the waiting
times between two consecutive bursts which are observed in the
intermittent state corresponding to $\sigma=260$. The results for
both modes are presented in Fig.~\ref{Fig: waiting T 2 unst}, where
we observe that the PDF of the waiting times for both modes is
dominated by a heavy-tail function with an exponent $3/2$.
Therefore, our results show that even in situations relatively far
from the instability onset, we can still observe stabilisation and
critical transition induced by pure additive noise. We note,
however, that the noise strength needs to be much larger now (up to
$\sigma=270$) to completely stabilise the dominant modes.

\section{Conclusions}
\label{Sec: conclusions} We have undertaken an analytical and
numerical study on the effect of additive noise on active nonlinear
SES. For simplicity we have focused on SES with quadratic
nonlinearities, in particular the KS equation, and randomly
perturbed in the vicinity of a primary bifurcation point, where the
dynamics is described by a single unstable dominant mode.

By adding a stochastic forcing acting on the first stable mode, we
have been able to provide a detailed and systematic investigation of
the transitions between different states. In particular, we have
observed that at low values of the noise strength, the amplitude of
the dominant mode is dominated by finite fluctuations around the
zero--noise solution ($\sigma=0$). On the other hand, at high values
of the noise strength the dominant mode can be completely
stabilised, yielding the zero solution when the spatio-temporal
evolution of the system is time averaged. These two states are
continuously connected by another intermediate state where the
solution intermittently fluctuates between the first non-zero state
and the final zero state, and it is characterised by a burst-like
dynamics. The transitions between the different states have been
completely characterised through critical exponents, obtaining
excellent agreement between theory and numerical results. In
addition, we have been able to rigorously derive the critical
exponent $3/2$ describing the waiting times of the intermittent
state II. It is important to emphasise that the derivation has been
done for general SES whose dominant mode amplitude  is given by
Eq.~(\ref{Eq:SL}). In this sense, the exponent $3/2$ reveals the
existence of an underlying universal mechanism dictated by the
random walk properties. Interestingly, the same exponent has been
ubiquitously found in many physical systems that display avalanche or
burst-like dynamics.  Examples include neuronal
avalanches in the spontaneous cortex
activity~\cite{neuronal_on_off}, on-off intermittency in
electroconvection of nematic liquid crystals~\cite{crystal_on_off},
or interface dynamics in disordered
media~\cite{imbibition_on_off_2,imbibition_on_off_1}.

When the noise acts on the whole subspace of stable modes (or
equivalently on the first and second stable modes) the dynamics of
the dominant mode is corrected by an additive noise which keeps the
system away of being completely stabilised, with the loss of any
critical intermittent behaviour. Again, the numerical results in this
situation are found to be in very good agreement with the analytical
predictions. We have also considered a case where the noise acts on
the second stable mode only (or equivalently on the stable odd
modes). In this situation the quadratic nonlinearity can filter out
the noise such that the dynamics of the  dominant mode is not
affected by the stochastic forcing applied to the system.

It is important to remark that, although the theory presented in
this work has been applied to the KS equation, our analysis is
rather general and could be easily extended to other models, such as
the stochastic Burgers equation~\cite{DBMHGP08, Ro:03} , used for example as a prototype for 1D
turbulence albeit without pressure gradient, the Kardar-Parisi-Zhang
(KPZ) equation, largely studied in the context of surface
growth~\cite{Bara1995}, or the stochastic Swift-Hohenberg equation,
often used as a model for Rayleigh-B\'enard convection, which in
turn is commonly used as a prototype to study instabilities out of
equilibrium in SES~\cite{Cross.etal_1993}. It should also be noted
that the approach followed in this study is very robust:
The stochastic SL equation (\ref{Eq:SL}) is
the universal amplitude equation for SPDEs with quadratic
nonlinearities, as proved in Ref.~\cite{Blomker.etal_2007}.
Furthermore, the characterisation of the different regimes depending
on the strength of the noise, and in particular the calculation of
the critical exponent that describes the intermittent behaviour are
systematic and rigorous. It is expected that the techniques and
methodologies developed in this study can be extended to more
general classes of problems, where noise induced phenomena occur.

Finally, by performing numerical integrations of the noisy KS equation in
a regime relatively far from the instability onset, where there are
two unstable dominant modes, we have observed similar stabilisation
and state transitions induced by an additive noise which is acting
on the first stable mode only. It is important to remark that,
although there is no theory in this case, our numerical study
provides evidence that such stabilisation and noise induced state
transitions are not restricted to a regime close to the instability
onset.
\appendix
\section{Derivation of the amplitude equation}\label{App amplitude eq}

We use techniques from homogenisation theory~\cite{PavlSt08} and singular perturbation theory for Markov
processes~\cite{Pap76} (see also~\cite{MTV01}) to
derive the amplitude equation that describes the dynamics of
Eq.~(\ref{Eq:Model scaled}) near the bifurcation point.

We consider a finite-dimensional truncation of Eqs.~(\ref{Eq:Scale
sep}) up to $N$ modes with a finite-dimensional null space,
i.e.~$\mbox{dim} (\mN) = N_0$, so that the number of fast modes is
$M=N-N_0$. The kernel $\mN$ of $\mL$ is then spanned by the first
$N_0$ eigenfunctions, and we can write:
$$
v_1 = \sum_{k=1}^{N_0} a_k \eb_k \quad \mbox{and} \quad v_\perp = \sum_{k=N_0+1}^{N} y_k \eb_k,
$$
which when introduced into Eqs.~(\ref{Eq:Scale sep}) yields the
following system of equations:
\begin{subequations}
\label{Eq:fast-slow}
\begin{align}
\dot{a}_m &=\frac{1}{\epsilon}f_0^{m}(\boldsymbol a,\boldsymbol y)+f_1^m(\boldsymbol a), \quad m=1, \dots N_0, {} \\
\dot{y}_k &=-\frac{1}{\epsilon^2}\lambda_k y_k+\frac{1}{\epsilon}g_0^k(\boldsymbol a,\boldsymbol y)+g_1^k(\boldsymbol y)+
\frac{1}{\epsilon}\sigma q_k\dot{\beta}_k,\quad k=N_0+1, \dots N
\end{align}
\end{subequations}
where $\boldsymbol a=(a_1,\dots,a_{N_0})^T$ and $\boldsymbol
y=(y_{N_0+1},\dots,y_{N})^T$.
The different vector field terms in the above system of equations are given as:
\begin{subequations}
\label{Eq:fast-slow terms}
\begin{align}
f_0^m(\boldsymbol a,\boldsymbol y)& = 2\sum_{n=1}^{N_0}a_n\sum_{\ell=N_0+1}^{N}B_{n\ell m}y_\ell
+\sum_{n,\ell=N_0+1}^{N}B_{n\ell m}y_ny_\ell, \label{Eq:fast-slow terms a} {} \\
g_0^k(\boldsymbol a,\boldsymbol y)& = \sum_{n,\ell=1}^{N_0}B_{n\ell k}a_n a_\ell
+2\sum_{n=1}^{N_0}a_n\sum_{\ell=N_0+1}^{N}B_{n\ell k}y_\ell+
\sum_{n,\ell=N_0+1}^{N} B_{n\ell k}y_n y_\ell, \label{Eq:fast-slow terms b} {} \\
f_1^m(\boldsymbol a)&= j_m a_m, \qquad g_1^k(\boldsymbol y)= j_k y_k,
\end{align}
\end{subequations}
where the coefficients $j_k$ and $\lambda_k$ are defined as  $\mJ\eb_k=j_k\eb_k$, and $-\mL\eb_k=\lambda_k\eb_k$,
respectively, and
\begin{equation}\label{Eq: B_nlk}
 B_{n\ell k}=\left(\mathcal{B}(\eb_n,\eb_\ell),\eb_k\right),
\end{equation}
with the inner product $(f,g)=\int_{-\pi}^{\pi}f(x)
g(x)\:\mathrm{d}x$.
In the above equations (\ref{Eq:fast-slow}) we have assumed the condition
$\mathcal{P}_c\mathcal{B}(\eb_m,\eb_n)=0$ for the null space, where $n,m=1,\dots,N_0$,
and we shall also assume the condition
$\mathcal{P}_c\mathcal{B}(\eb_k,\eb_k)=0$ for the bilinear map, so
that we have
\begin{equation}\label{e:centeringI}
B_{kkm}=0, \quad m=1,\dots,N_0.
\end{equation}
This assumption, which is satisfied for the Burgers nonlinearity $u \partial_{x}u$, ensures that the centering condition from homogenisation theory, see Eq.~\eqref{e:centering} below, is satisfied. The system of
equations (\ref{Eq:fast-slow}) for $a_m$ and $y_k$ is of the form of
a fast-slow system of stochastic differential equations for which
the associated backward-Kolmogorov equation (for $w^{\epsilon} = \mathbb{E} (f(\boldsymbol{a}(t), \boldsymbol{y}(t) |\boldsymbol{a}(0)=\boldsymbol{a}, \boldsymbol{y}(0) =  \boldsymbol{y})$) reads:
\begin{equation}
\partial_t w^{\epsilon}=(\epsilon^{-2}\mL_0+\epsilon^{-1}\mL_1+
\mL_2)w^{\epsilon}, \label{BackK}
\end{equation}
where
\begin{subequations}
\begin{align*}
\mL_0 =& \sum_{k=N_0+1}^{N}\bigg(-\lambda_k y_k\partial_k+\frac{\sigma^2 q_k^2}{2}\partial_k^2\bigg), {} \\
\mL_1 =& \sum_{m=1}^{N_0}f_0^m(\boldsymbol a,\boldsymbol y)\partial_m+ \sum_{k=N_0+1}^{N}g_0^k(\boldsymbol a,\boldsymbol y)\partial_k, {} \\
\mL_2 = & \sum_{m=1}^{N_0}f_1^m(\boldsymbol a)\partial_m+\sum_{k=N_0+1}^{N}g_0^k(\boldsymbol y)\partial_k,
\end{align*}
\end{subequations}
and $\partial_m$ and $\partial_k$ represent derivatives respect to $a_m$ and $y_k$ for
$m=1,\dots,N_0$ and $k=N_0+1,\dots,N$, respectively.
Here, the operator $\mL_0$ corresponds to the generator of a finite-dimensional Ornstein-Uhlenbeck
(OU) process, so that the invariant measure of the fast process is Gaussian:
\begin{equation}
\rho(d\boldsymbol
y)=\frac{1}{\mathcal{Z}}\exp\Bigl({-\sum_{k=N_0+1}^{N}\frac{\lambda_k}{\sigma^2q_k^2}y_k^2}\Bigr)\mathrm{d}\boldsymbol
y,
\end{equation}
where $\mathcal{Z}$ is the normalisation constant.
From~\eqref{Eq:fast-slow terms a} and Assumption~\eqref{e:centeringI} we deduce that the vector field $f_0^m(\boldsymbol a,\boldsymbol y)$
is centered with respect to the invariant measure of the fast process,
\begin{equation}\label{e:centering}
\int_{\R^M} f_0^m(\boldsymbol a ,\boldsymbol y) \, \rho (d \boldsymbol y) = 0.
\end{equation}
We will now show that the amplitude equation is of the form
\begin{equation}\label{e:homog_M}
d \boldsymbol z = \bar{\boldsymbol v}_M(\boldsymbol z) \, dt + \bar{\boldsymbol g}_M(\boldsymbol z) \, d W,
\end{equation}
where $ W$ denotes a standard $N_0-$dimensional Wiener process. The
subscript $M$ used here is to emphasise the fact that the
homogenised coefficients depend on the number of fast processes that
we take into account. The calculation of the coefficients
$\bar{\boldsymbol v}_M(\boldsymbol z)$ and $\bar{\boldsymbol g}_M(
\boldsymbol z)$ which appear in Eq.~(\ref{e:homog_M}) will
be obtained by using homogenisation theory.

We solve the backward-Kolmogorov equation~\eqref{BackK} by looking for a solution  in the form of a power series expansion in $\epsilon$:
\begin{equation}
 w^{\epsilon}=w_0+\epsilon w_1+\epsilon^2 w_2+\mathcal{O}(\epsilon^3).
\end{equation}
Substituting this expansion into~(\ref{BackK}) and equating
coefficients of powers of $\epsilon$ to zero, we find:
\begin{subequations}
\begin{align}
\mL_0w_0& =  0, {} \label{Eq:Syst w0}\\
\mL_0w_1+\mL_1w_0 & = 0, {} \label{Eq:Syst w1}  \\
\mL_0w_2+\mL_1w_1+\mL_2w_0 & = \partial_t w_0. \label{Eq:Syst w2}
\end{align}
\end{subequations}
Since the null space of the generator of the OU process consists of constant in $\boldsymbol y$, from Eq.~(\ref{Eq:Syst w0}) we deduce that
$w_0=w_0(\boldsymbol a,t)$. The solvability condition for the second equation reads
\begin{equation}\label{Eq:solv 1}
\int_{\R^M} (\mL_1w_0) \, \rho (d \boldsymbol y) = 0.
\end{equation}
This solvability condition is satisfied on account of~\eqref{e:centering}, and the
fact that the first term in the expansion is independent of ${\boldsymbol y}$.
We now solve this equation by using the following ansatz:
\begin{equation}
 w_1= \biggl(a_m\sum_{\ell=N_0+1}^{N}\chi_{\ell m}y_\ell+
\sum_{n,\ell=N_0+1}^{N}\zeta_{n\ell m}y_ny_\ell\biggr)\partial_m w_0,
\end{equation}
where the constants $\chi_{\ell m}$ and $\zeta_{n\ell m}$ are to be
determined from Eq.~(\ref{Eq:Syst w1}) together with (\ref{Eq:solv 1}).
In particular,  we  obtain:
\begin{equation}
\chi_{k m}=\frac{2B_{mk m}}{\lambda_k}, \qquad \zeta_{nk m}=\frac{B_{nk m}}{\lambda_n+\lambda_k},
\end{equation}
for $m=1,\dots,N_0$ and $n,k=N_0+1,\dots,N$.
Finally, for Eq.~(\ref{Eq:Syst w2}) we have the solvability condition:
\begin{equation}\label{Eq:solv 2}
\int_{\R^M} \bigg(\partial_t w_0-\mL_2w_0-\mL_1 w_1\bigg) \, \rho (d \boldsymbol y) = 0,
\end{equation}
which allows us to obtain the homogenised SDE (\ref{e:homog_M})\footnote{To be more precise,
it allows us to obtain the homogenised backward Kolmogorov equation from which we can read
off the limiting SDE.}.
For the components of the drift term we get:
\begin{equation}\label{eq:drift}
v_M^m(\boldsymbol a)=\bigg\langle\biggl(f_1^m+a_m\sum_{k=N_0+1}^{N}g_0^k\chi_{km}+
2\sum_{n,k=N_0+1}^{N}g_0^ky_n\zeta_{nkm}+f_0^m\sum_{k=N_0+1}^{N}\chi_{km}y_k\biggr)\bigg\rangle,
\end{equation}
for $m=1,\dots,N_0$, and where $\langle\dots\rangle$ denotes average with respect to the invariant
measure $\rho(\mathrm{d}\boldsymbol y)$. Similarly, the components of the
quadratic form associated with the diffusion matrix $\bar{\boldsymbol g}_M^2$ are given by:
\begin{equation}\label{eq:diffusion}
\frac12(\bar{g}_M^2)_{ij}=\bigg\langle f_0^i\biggl(a_j\sum_{n=N_0+1}^N\chi_{nj}y_n+
\sum_{n,\ell=N_0+1}^N\zeta_{n\ell j}y_ny_\ell\biggr)\bigg\rangle,
\end{equation}
where $i,j=1,\dots,N_0$. The integrals in (\ref{eq:drift}) and (\ref{eq:diffusion})
are Gaussian integrals that can be calculated analytically. In particular, substitution of
$f_0^i$ and $g_0^i$ given in Eq.~(\ref{Eq:fast-slow terms a}) and
(\ref{Eq:fast-slow terms b}), respectively, into (\ref{eq:drift}) and
(\ref{eq:diffusion}), and calculation of the resulting Gaussian integrals lead to the
amplitude equation (\ref{e:homog_M}) which can be written as the following
system of coupled SL equations:
\begin{equation}\label{Eq:Syst SL}
 d a_m = v_M^m(a_1,\dots,a_{N_0})d t+\sum_{n=1}^{N_0}(\bar{g}_M)_{mn}d W_n,
\end{equation}
for  $m=1,\dots, N_0$, and where the different terms are given as:
\begin{subequations}
\begin{align}
v_M^m &=\bigg(j_m+\sum_{n,k=N_0+1}^{N}\!\!\!\frac{2B_{kmm}B_{nnk}}{\lambda_n\lambda_k}\sigma^2q_n^2\bigg)a_m+
\sum_{\ell=1}^{N_0}a_\ell\!\!\sum_{n,k=N_0+1}^{N}\frac{2B_{nkm}B_{\ell nk}}{\lambda_n(\lambda_n+\lambda_k)}\sigma^2 q_n^2 \nonumber{}\\
&- a_m\sum_{n,\ell=1}^{N_0}\!\! a_n a_\ell\!\!\sum_{k=N_0+1}^{N}\!\!\frac{2B_{kmm} B_{n\ell k}}{\lambda_k}, {} \\
\frac12(\bar{g}_M^2)_{ij}&=  a_ia_j\sum_{\ell=1}^{M}\frac{4 B_{\ell ii}B_{jnn}}{\lambda_\ell^2}\sigma^2q_\ell^2+\sum_{k,n=N_0+1}^{N}\frac{B_{kni}B_{knj}}{\lambda_k(\lambda_n+\lambda_k)^2}\sigma^4q_n^2q_k^2.
\end{align}
\end{subequations}
In the case of a one-dimensional null space
($N_0=1$) we define $A=a_1$, and the above expressions have a simpler form.
In particular we have:
\begin{subequations}
 \label{Eq:GL Ap}
\begin{align}
\bar{v}_M(A) &= (j_1+\gamma_1\sigma^2) A-\gamma_2 A^3, {} \\
\bar{g}_M(A) &= \sigma\sqrt{\gamma_a\sigma^2+\gamma_m A^2},
\end{align}
\end{subequations}
with the different coefficients given as:
\begin{subequations}
 \label{Eq:coefficients SL App}
\begin{align}
\gamma_1&=\sum_{n=2}^{N}\frac{2 B_{n11}^2}{\lambda_n^2}q_n^2+\sum_{n,\ell=2}^{N}\frac{2 B_{n\ell 1}B_{n1k}}{(\lambda_n+\lambda_\ell)\lambda_n}q_n^2
+\sum_{n,\ell=2}^{N}\frac{B_{n11}B_{nn\ell}}{\lambda_n\lambda_\ell}q_\ell^2, {} \\
\gamma_2&=\sum_{n=2}^{N}\frac{2 B_{n11}B_{11n}}{\lambda_n}, \quad \gamma_a =\sum_{n,\ell=2}^{N}\frac{2 B_{n\ell 1}^2}{(\lambda_n+\lambda_\ell)^2\lambda_\ell}q_n^2q_\ell^2 ,
\quad \gamma_m =\sum_{n=2}^{N}\frac{4 B_{n11}^2}{\lambda_n^2}q_n^2,
\end{align}
\end{subequations}
where the noise in the resulting amplitude equation is interpreted
in the It\^{o} sense.  We note that the form of the amplitude
equation (\ref{Eq:GL Ap}) does not depend on whether the noise is
interpreted as It\^{o} or Stratonovich, and the only difference
between both interpretations is the first term on the right hand
side of the expression for the coefficient $\gamma_1$, which exactly
corresponds to $\gamma_m/2$, and is a consequence of the
multiplicative term of the noise. Let us also remark that the
Stratonovich interpretation turns out to be more convenient for the
results we present in \S~\ref{Sec: Case study} for the noisy KS
equation, specially for the noise-induced state transitions.
Therefore, we will consider the amplitude equation (\ref{Eq:GL Ap})
with the coefficients given by
 Eq.~(\ref{Eq:coefficients SL App}) with $\gamma_1\to \gamma_1-\gamma_m/2$.

It is also interesting to consider the case  when the null space is two-dimensional. By
assuming the Burgers nonlinearity ($u\partial_xu$), we can explicitly work out the
different terms of the amplitude equation (\ref{Eq:Syst SL}) for the two
components  $a_1(t)$ and $a_2(t)$, obtaining:
\begin{subequations}
\label{Eq:Syst SL 2D}
\begin{eqnarray}
v_M^1  &=& - \frac{1}{4 \lambda_2}a_1^3 - \frac{1}{4 \lambda_2} a_1 a_2^2 +
\big( j_1  + V_M\big)a_1, \\ 
v_M^2  &=& - \frac{1}{4 \lambda_2}a_2^3 - \frac{1}{4 \lambda_2} a_1^2 a_2 +
\big( j_2  + V_M\big)a_2,\\ 
\frac12(\bar{g}_M^2) & = &    \left(
\begin{array}{cc}
 \frac{\sigma^2}{8}\,\frac { r_2^2 a_2^2
}{\lambda_2^2}+ \frac{\sigma^2}{8}\,\frac{q_2^2 a_1^2}{\lambda_2^2} +G_M
 \; & \;  \frac{\sigma^2}{8}\,\frac {\left( r_2^{2}-
 q_2^{2} \right)}{\lambda_2^2} a_1 a_2
\\
  \frac{\sigma^2}{8}\,\frac {\left( r_2^{2}-
 q_2^{2} \right)}{\lambda_2^2} a_1 a_2 \;  & \;  \frac{\sigma^2}{8}\,\frac { r_2^2 a_1^2
}{\lambda_2^2}+ \frac{\sigma^2}{8}\,\frac{q_2^2 a_2^2}{\lambda_2^2}+ G_M
\end{array}
\right) ,
\end{eqnarray}
\end{subequations}
with the coefficients:
\begin{subequations}
\begin{eqnarray*}
V_M&=&\frac{\sigma^2}{8}\sum_{k=2}^{M}\frac{k \lambda_k (q_{k+1}^2 +
r_{k+1}^2) -\lambda_{k+1} (q_k^2 +r_k^2) (k+1)}{(\lambda_{k+1} + \lambda_k) \lambda_k \lambda_{k+1}}, \\
G_M &=& \frac{\sigma^4}{16} \,\sum _{k=2}^{M} \frac {  q_k^2 q_{k+1}^2+ r_k^2 r_{k+1}^2}
{\lambda_k ( \lambda_{k+1} +\lambda_k ) \lambda_{k+1} },
\end{eqnarray*}
\end{subequations}
where we have assumed a general case where the two components of the noise for each fast
mode (which arise as a consequence of the fact that the null space is multidimensional) have different
wave number dependence, namely $q_k$ and $r_k$. We therefore see that in general,
the diffusion matrix is non diagonal and  the resulting amplitude equations are coupled not
only through the drift  term but also through the noise term.

\section{Derivation of the noisy Kuramoto--Sivashinsky equation for
a hydrodynamic system}\label{App noisy KS}

We provide a physical example where the noisy KS
equation can be derived from a general formulation in the framework
of thin-film hydrodynamics.

\begin{figure}
\centering
\includegraphics[width=0.75\textwidth]{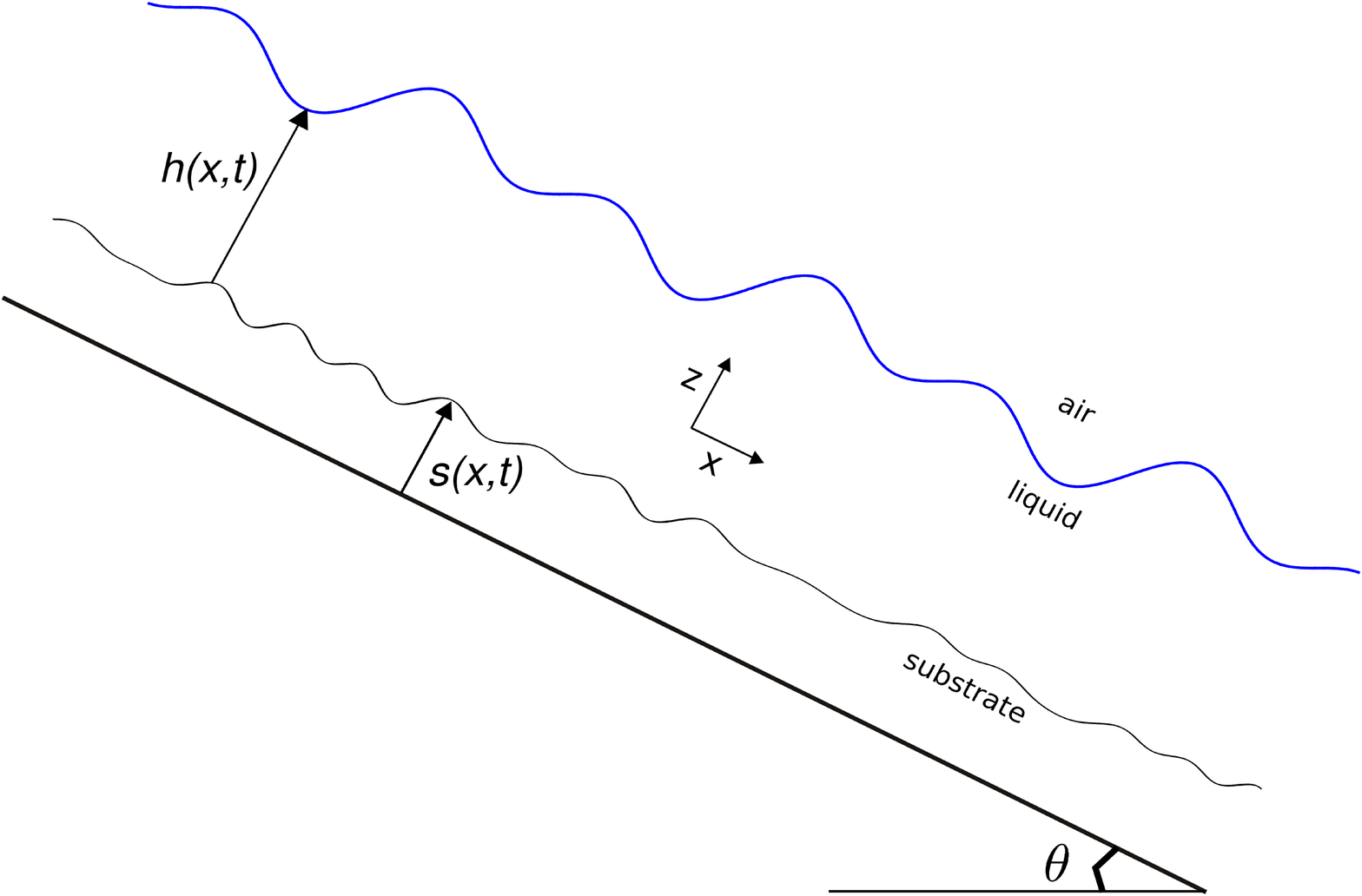}
\caption{Sketch of a liquid film falling down over an inclined plane
with a vibrating disordered wall. The thickness of the film and
the disordered wall position are  denoted as $h(x,t)$ and $s(x.t)$,
respectively.}\label{Fig: setup}
\end{figure}

Consider gravity-driven thin film flow down an uneven wall that is
inclined at an angle $\theta$ to the horizontal, as shown in
Fig.~\ref{Fig: setup}. Let $x$ and $z$ be the stream- and the
cross-stream coordinates, respectively, and the wall be described by
$z=s(x,t)$ relative to a datum plane and where the time-dependence
is such that each point of the wall fluctuates only in the
$z$-direction, i.e., at a given time $t$ the velocity of the wall
point $(x, s(x, t))$ is the vector $(0, s_t(x, t))$. Let also the
free surface be located at $z=f(x, t)$. We introduce dimensionless
variables by utilising the Nusselt film thickness $h_0$
(corresponding to undisturbed flow down an inclined plane) as a
lengthscale, the Nusselt surface velocity
$U_0=\rho\mathrm{g}h_0^2\sin\theta/2\mu$, where $\rho$ and $\mu$ are
the density and the dynamic viscosity of the liquid, respectively,
and $\mathrm{g}$ is gravity, as a velocity scale and $U_0/h_0$ as a
time scale. The pressure scale is chosen as $\mu U_0/h_0$. We
additionally introduce the Reynolds number $Re=\rho U_0 h_0/\mu$ and
the capillary number $Ca=\mu U_0/\gamma$
where $\gamma$ denotes the surface tension.

The governing equations are given by the incompressible Navier--Stokes equations:
\begin{eqnarray}
&&Re(u_t+uu_x+vu_z)=-p_x+u_{xx}+u_{zz}+2,\\
&&Re(v_t+uv_x+vv_z)=-p_z+v_{xx}+v_{zz}-2\cot\theta,\\
&&u_x+v_z=0,
\end{eqnarray}
where subindices represent partial derivatives, $u$ and $v$ denote
the $x$- and the $z$-component of the velocity in the liquid,
respectively, and $p$ denotes the deviation of the pressure from the
atmospheric level. The boundary conditions at the wall, $z=s(x,
t)$, are given by:
\begin{equation}
u=0,\qquad v=s_t(x, t),
\end{equation}
and at the free surface, $z=f(x, t)$, the kinematic compatibility and the tangential
and normal stress balance conditions are satisfied:
\begin{eqnarray}
&& f_t+uf_x-v=0,\\
&& (1-f_x^2)(u_z+v_x)+2f_x(v_z-u_x)=0,\\
&& p=\frac{2}{1+f_x^2}(-f_x(u_z+v_x) +u_xf_x^2+v_z)-\frac{f_{xx}}{Ca(1+f_x^2)^{3/2}}.
\end{eqnarray}
We consider here the long-wave approximation so that we  introduce
the so-called thin-film or long-wave parameter $\varepsilon$,
defined as the ratio of the typical film thickness to the length
scale over which streamwise variations occur. We then introduce new
variables: $\xi=\varepsilon x$, $\tau=\varepsilon t$, and
$w=v/\varepsilon$. We also assume that $Ca=O(\varepsilon^2)$ and
define $\widetilde{C}a=Ca/\varepsilon^2$.
%
We expand the different variables as, $u=u_0+\varepsilon
u_1+\cdots$, $w=w_0+\varepsilon w_1+\cdots$, and $p=p_0+\varepsilon
p_1+\cdots$, which lead to a series of solvable perturbation
problems. The leading-order problem is:
\begin{equation}
u_{0zz}=0,\qquad p_{0z}=-2\cot\theta,\qquad u_{0\xi}+w_{0z}=0.
\end{equation}
At $z=s(\xi, \tau)$,
$
u_0=0,\, w_0=s_\tau(\xi, \tau).
$ 
At the free surface, $z=f(\xi, \tau)$, the tangential and normal stress balance conditions imply
$ 
u_{0z}=0, \, p_0=-f_{\xi\xi}/\widetilde{C}a.
$ 
The solution of the problem at leading order is:
\begin{eqnarray}
&&u_0=-(z-h-s)^2+h^2,\label{eq:u0}\\
&&w_0=u_0(h+s)_\xi-[h^2]_\xi(z-s)+s_\tau(\xi, \tau),\label{eq:w0}\\
&&p_0=-2(\cot\theta)(z-h-s)-(h+s)_{\xi\xi}/\widetilde{C}a,\label{eq:p0}
\end{eqnarray}
where we introduced for convenience the film thickness $h=f-s$.
Note that the kinematic compatibility condition can be written as
\begin{equation}
h_\tau+q_\xi=0,
\end{equation}
where $q=\int_s^fu\:\mathrm{d}z$. Using (\ref{eq:u0}), we find
$
q=\frac{2}{3}h^3+O(\varepsilon),
$ 
which implies
\begin{equation}\label{eq:LongWave_O_0}
\textstyle h_\tau+\bigl[\frac{2}{3}h^3\bigr]_\xi+O(\varepsilon)=0.
\end{equation}
%
At next order, we obtain the following system of equations:
\begin{eqnarray}
&&u_{1zz}=Re(u_{0\tau}+u_0u_{0\xi}+w_0u_{0z})+p_{0\xi},\\
&&p_{1z}=w_{0z},\\
&&u_{1\xi}+w_{1z}=0,
\end{eqnarray}
subject to
$
u_1=w_1=0
$ 
at $z=s$ and
$
u_{1z}=0,\,
p_1=2w_{0z}-2f_\xi u_{0z}
$ 
at $z=f$, where $u_0$, $w_0$, and $p_0$ are given by
(\ref{eq:u0})--(\ref{eq:p0}). The time derivative $u_{0\tau}$
involves the time derivative $h_\tau$, as is evident from
(\ref{eq:u0}). The latter is eliminated by using
(\ref{eq:LongWave_O_0}). For brevity we do not show the solution of
the problem at first order (which can be easily found with a
symbolic manipulation software) as it turns out to be rather
lengthy. Substituting $u=u_0+\varepsilon u_1+O(\varepsilon^2)$ into
$q=\int_s^fu\:\mathrm{d}z$, we find
\[q=\frac{2}{3}h^3+\varepsilon h^3\biggl[\frac{8Re}{15}h^3h_\xi-
\frac{2\cot\theta}{3}(h+s)_\xi+\frac{1}{3\widetilde{C}a}(h+s)_{\xi\xi\xi}\biggr]
+O(\varepsilon^2). \]
Thus, we obtain the following evolution equation for the 
film thickness, $h$:
\begin{equation}
h_\tau+\biggl(\frac{2}{3}h^3+\varepsilon
h^3\biggl[\frac{8Re}{15}h^3h_\xi-
\frac{2\cot\theta}{3}(h+s)_\xi+\frac{1}{3\widetilde{C}a}(h+s)_{\xi\xi\xi}\biggr]\biggr)_\xi=0.
\end{equation}
We proceed next with a weakly nonlinear analysis. We assume that
both the amplitude of the free surface and the bottom wall are
small, of $O(\varepsilon)$, and we write $h=1+\varepsilon\eta$ and
$s=\varepsilon\sigma$. Substituting these expressions into the
evolution equation above re-written in the moving frame,
$\chi=\xi-2\tau$,
we obtain the following evolution equation: 
\begin{equation}
\bar{\eta}_{\tilde{\tau}}+
4\bar{\eta}\bar{\eta}_\chi+D\bar{\eta}_{\chi\chi}+\frac{1}{3\widetilde{C}a}\bar{\eta}_{\chi\chi\chi\chi}
=\Sigma,
\end{equation}
where $O(\varepsilon^2)$ terms have been neglected,
$\tilde{\tau}=\varepsilon\tau$,
$\bar{\eta}(\chi, \tilde{\tau})=\eta(\chi+2\tilde{\tau}/\varepsilon, \tilde{\tau}/\varepsilon)$,
$\bar{\sigma}(\chi, \tilde{\tau})=\sigma(\chi+2\tilde{\tau}/\varepsilon, \tilde{\tau}/\varepsilon)$,  
$D = 8Re/15-2\cot\theta/3$ and
$\Sigma = (2\cot\theta/3)\bar{\sigma}_{\chi\chi}-(1/3\widetilde{C}a)\bar{\sigma}_{\chi\chi\chi\chi}$.
%
We find that $s(\xi, \tau)=\epsilon\bar{\sigma}(\xi-2\tau,
\epsilon\tau)$, which physically means that the typical deformation
of the topography shape is of small amplitude and large wavelength
(of the same order as the wavelength of the typical free-surface
wave), it propagates downstream with constant velocity, and is
slowly changing in the
frame moving with this velocity. 

To simplify the latter evolution equation, we introduce the
transformation, $\bar{\eta}=U/A$, $\chi=X/B$, and
$\tilde{\tau}=T/C$, where $A=4/(3\widetilde{C}a|D|^3)^{1/2}$,
$B=(3\widetilde{C}a|D|)^{1/2}$, and $C=3\widetilde{C}aD^2$, which
leads to the following canonical form:
\begin{equation}
U_T+UU_X\pm U_{XX}+U_{XXXX}=S(X, T),
\end{equation}
where $S=(A/C)\Sigma$, and the sign $+$/$-$ corresponds to
the positive/negative value of $D$. We therefore obtain the noisy KS
equation (\ref{Eq:KS}).\\


{\bf Acknowledgements.} We acknowledge financial support from EU-FP7
ITN Grant No. 214919 (Multiflow), ERC Advanced Grant No. 247031 and the EPSRC Grant No. EP/H034587/1.
The work of DTP was also supported in part by the National Science Foundation grant DMS-0707339.

\end{document}